\documentclass[12pt,a4paper]{article}
\usepackage{amssymb} 
\usepackage{amsmath} 
\usepackage{xcolor} 
\usepackage[compat=1.1.0]{tikz-feynman} 
\usepackage{braket} 

\usepackage{caption} 
\usepackage{subcaption}

\usepackage[utf8]{inputenc}
\usepackage[margin=0.8 in]{geometry}    
\usepackage[title,titletoc,toc]{appendix}
\usepackage{hyperref}     
 \usepackage{color}
 \usepackage[most]{tcolorbox}
 \tcbset{colback=yellow!10!white, colframe=red!50!black, 
        highlight math style= {enhanced, 
            colframe=red,colback=red!10!white,boxsep=0pt}
        }        
\usepackage[parfill]{parskip}    
\usepackage{graphicx}
\usepackage{amsfonts}

\title{A Finite Energy-Momentum Tensor for the $\phi^3$ theory in $6$ dimensions}
\author{ Pavan Dharanipragada and B. Sathiapalan\\Institute of Mathematical Sciences\\
CIT Campus, Tharamani\\ Chennai 600113, India\\
and\\ Homi Bhabha National Institute\\Training School Complex, Anushakti Nagar\\
Mumbai 400085, India\\pavand, bala@imsc.res.in}

\begin{document}
	\begin{flushright}
		IMSc/2021/05/02
	\end{flushright}
{\let\newpage\relax\maketitle}

\maketitle
\begin{abstract}
  Following Brown\cite{brown_dimensional_1980}, we construct composite operators for the scalar $\phi^3$ theory in six dimensions using renormalisation group methods with dimensional regularisation. We express bare scalar operators in terms of renormalised composite operators of low dimension, then do this with traceless tensor operators. We then express the bare energy momentum tensor in terms of the renormalised composite operators, with some terms having divergent coefficients. We subtract these away and obtain a manifestly finite energy tensor. The subtracted terms are transverse, so this does not affect the conservation of the  energy momentum tensor. The trace of this finite \emph{improved} energy momentum tensor 
  vanishes at the fixed point indicating conformal invariance. Interestingly it is not RG-invariant except at the fixed point, but can be made RG invariant everywhere by further addition of transverse terms, whose coefficients vanish at the fixed point.

\end{abstract}
\thispagestyle{empty}
\newpage
\tableofcontents
\thispagestyle{empty}
\newpage
\setcounter{page}{1}




\newcommand{\pole}[2][]{\frac{#2}{#1 64\pi^3 (d-6)}}
\newcommand{\gder}[1]{\frac{\partial #1}{\partial g}}
\newcommand{\mud}{\mu^{d/2-3}}
\newcommand{\flow}[1]{\mu\frac{\mathrm d #1 }{\mathrm d\mu}}
\newcommand{\dd}{\mathrm d}
\newcommand{\pr}{^\prime}
\newcommand{\gmn}{g_{\mu\nu}}
\newcommand{\tmn}{t_{\mu\nu}}
\newcommand{\smn}{s_{\mu\nu}}
\newcommand{\emt}{T_{\mu\nu}}
\newcommand{\nemt}{\Theta_{\mu\nu}}
\newcommand{\psq}{\partial^2}



\section{Introduction}
The advent of AdS/CFT and holography has made the study of field theories in various dimensions more interesting. Conformal field theories  can be expected to have gravity duals in AdS space but more generally even non-conformal theories can be expected to have duals because they can be viewed, at least in principle as arising from an RG flow of a conformal field theory perturbed by some non-marginal operator.
One such field theory is the $O(N)$ $\phi^4$ model in $4>d>2$, which been studied in the context of condensed matter physics and also subsequently in the AdS/CFT context.

Apart from applications, the $\phi^4$ field theory in $d=4$ has been studied just as a model field theory to develop and explore new techniques. $\phi^4$ field theory exists as a non-trivial interacting field theory in $d<4$. Non-trivial fixed points are known\textemdash the ``Wilson-Fisher" fixed points\textemdash and these have been much studied. In $d=4$, in the absence of a non-trivial fixed point, the interacting theory is thus non perturbatively not renormalizable (defined). Nevertheless it is a useful place to explore perturbative issues.  A famous example of this is Polchinski's proof of perturbative renormalizability of this theory in $d=4$ using the Wilsonian Exact Renormalization Group ideas\cite{polchinski_renormalization_1984}.

Another example is the study of the composite operators in this theory ($\phi^4$ in $d=4$) by Brown\cite{brown_dimensional_1980}. Composite operators (also called normal products in older literature) have to be renormalised after systematically subtracting divergences resulting from products of fields and their derivatives taken at a single spacetime point\cite{zimmermann_local}. Since various currents in the theory are composite operators, their precise definition requires well-defined renormalised composite operators. Techniques such as the \emph{action principle}\footnote{The action principle\cite{Breitenlohner:1977hr} specifies the change of correlation functions as the parameters or fields in the theory are changed. Thus it provides an easy way to derive various identitites and other structural properties. The manipulations in sections \ref{phi2}, \ref{phi3} make use of this principle.} have been used to make the renormalisation easier, especially in dimensional regularisation\cite{collins_normal_1975}. These techniques were applied to quantise currents. Brown applies these techniques to $\phi^4$ theory and derives a manifestly finite energy momentum tensor for the theory. These and similar ideas and techniques have been fruitfully applied to Yang-Mills theories\cite{KlubergStern:1974rs,Zee:1983mj,Freeman:1983cx,Dudal:2008tg,Spiridonov:1984br}, QED\cite{Hathrell:1981gz}, non-linear $\sigma$ models\cite{Osborn:1987au}, and others\cite{Benedetti:2020yvb,Lalak:2018bow,Bellucci:2003ud,Mauri:2021ili}.

``Improving" energy momentum tensors has been the focus of much study since the early papers of Coleman, Jackiw, Callan (CCJ)\cite{callan_new_1970, Lowenstein:1971vf, Freedman:1974gs, Freedman:1974ze, collins_renormalization_1976,Polchinski:1987dy}. The EM tensor defined canonically is in general divergent, and not unique. But adding a transverse term (a term with the derivative $\partial_\mu\partial_\nu-g_{\mu\nu}\partial^2$) does not affect the Poincar\'e generators, while modifying the dilatation and conformal generators\cite{Coleman:1970je}. It is desirable to choose a finite EM tensor as its matrix elements  play a role in describing scattering in the weak gravitational field regime\cite{callan_new_1970,Christensen:1977jc}. Furthermore, in the context of AdS/CFT the energy momentum tensor of the boundary theory is the source for the bulk graviton.
That the divergent terms in the EM tensor can be cancelled by adding such transverse terms is a consequence of the fact that the EM tensor must be conserved and hence it can only include terms which are proportional to the equation of motion operators\cite{brown_dimensional_1980}, which are finite and vanish on-shell, and transverse terms. Hence only the latter terms could be divergent.

There is a certain ambiguity in the definition of the EM tensor given that any finite transverse term could be added to it without affecting the properties of the Poincar\'e generators. The improvement term coefficient $1/6$ by CCJ for $\phi^4$ theory is one such choice, (the desirability of this term for finite temperature correlation functions is shown in \cite{Bibilashvili:1992hx}). Freedman and Weinberg\cite{Freedman:1974ze} show that this is in general does not renormalise EM tensor order by order in perturbation theory beyond three loops. Collins\cite{collins_renormalization_1976} shows that the CCJ improvement term indeed renormalises the EM tensor, but only non-perturbatively, and that order by order, it needs to be modified by terms with coefficients which are powers of $(d-4)$. He further shows that this is the unique improvement term that fulfills certain criteria for the EM tensor, one of which is that the EM tensor must be RG invariant, i.e., $\flow{\emt}=0$. Joglekar and Misra\cite{Joglekar:1988uc} extend this uniqueness proof to include improvement terms that are \emph{finite functions of bare quantities}. But it is not clear a priori that RG invariance of the EM tensor is a required property. If, for instance, it can be related to the RG invariance of the S-matrix for graviton scattering then that would be an argument for it.                                            

We follow Brown in renormalising the EM tensor for $\phi^3$ theory in six spacetime dimensions, by \emph{minimal subtraction}. The ``improved" energy tensor is constructed whose trace is proportional to the beta functions of the theory and consequently vanishes at fixed points (on using the equations of motion) for $d\neq 6$ as well. Vanishing trace condition is important because it can be taken as the definition of conformal invariance. The EM tensor thus plays a central role in the study of conformal field theories. The RG derivative of the EM tensor vanishes at the fixed point as well, as it should to preserve consistency. We also add a finite improvement term to obtain an RG-invariant, finite EM tensor. 

As mentioned, $\phi^3$ theory in six dimensions is the focus of this paper. This theory has the added advantage of being asymptotically free and thus has many features found in QCD\cite{Cardy:1974af,Lovelace:1974mi,Cardy:1975fz,Ma:1975vn,Cornwall:1995dr}. This has been one motivation for its study\cite{collins_1984,srednicki_2007,Toms:1982af}.
It is also related to the Yang-Lee edge singularity \cite{fisher_yang-lee_1978,Gracey:2015tta,deAlcantaraBonfim:1981sy,An:2016lni}. Furthermore, in $6-\epsilon$ dimensions, fixed points analogous to the Wilson Fisher fixed points can be found\cite{Fei:2014yja}. These are thus conformal field theories\cite{Mack:1973kaa,Gliozzi:2014jsa}. In $d=6$  asymptotic freedom would normally imply that these theories can be defined non-perturbatively. However the nature of the potential -$\phi^3$- makes the perturbative vacuum unstable and once again non-perturbatively these theories are ill-defined. Nevertheless, these pathologies don't show up in perturbation, and the potential can be studied in perturbation theory.


This paper is organized as follows. In Section \ref{thetheory} we give some background about the $\phi^3$ model. The counterterms and renormalisation group coefficients are defined in dimensional regularisation. In section \ref{compositeOperators} we use some properties of the renormalisation group to derive relations between the bare and renormalised composite scalar operators\cite{brown_dimensional_1980}. The main idea is to require finiteness of correlation functions with composite operator insertions in the renormalized theory. This involves the \emph{mixing} of higher dimension operators with all operators of lower dimension. Thus, $\phi^2$ mixes with $\phi$ and $\psq\phi$, while $\phi^3$ mixes with $\phi^2$, $\phi$, $\psq\phi$, $\psq\phi^2$, $\phi\psq\phi$ and $\partial^4\phi$. The bare and renormalised operators are arranged in two column vectors and their mixing is expressed as a matrix equation, with the coefficients arranged in a $7\times7$ matrix. Some of the matrix elements remain undetermined. We define an \emph{anomalous dimension matrix} that helps us express the undetermined quantities in terms of new anomalous dimensions associated with specific operator mixings. In section \ref{tensorOperators}, we use the results obtained in section \ref{compositeOperators} to renormalise traceless tensor operators with two indices. Finally in section \ref{emtensor},this allows us to separate the divergent terms in the canonical EM tensor, allowing us to write down an improvement term added to the canonical EM tensor which just subtracts the divergences.


We have added two appendices that involve Feynman diagram computations for counterterms and RG functions of the theory to one-loop in appendix \ref{appendix:oneloop}, and for the renormalisation $\phi^2$ and $\phi^3$ operators to one-loop level in appendix \ref{oneloopreno}. The former has previously been done to higher orders\cite{macfarlane_3_1974,Gracey:2015tta} and the latter for $\phi^2$ to one-loop\cite{collins_1984}, but they have been included here in an effort to keep this paper self-contained. The results in \ref{oneloopreno} serve as a check to the computations in \ref{compositeOperators}.

\section{$\phi^3$ theory in $6$ dimensions}
\label{thetheory}
The bare Lagrangian is
\begin{equation}
\mathcal{L}=\frac{1}{2}(\partial_{\mu}\phi_0)^2-\frac{1}{2}m_0^2\phi_0^2-\frac{1}{6}g\phi_0^3 +j_0\phi_0
\label{lagrangian}.
\end{equation}
The coupling $g$ is marginal in $6$ dimensions. We have had to include the linear term because the theory lacks $Z_2$ symmetry.

Splitting this into renormalised and counterterm lagrangian, 
\begin{equation}
\mathcal{L}=\mathcal{L}_R+\mathcal{L}_{CT},
\end{equation}
\begin{equation}
\mathcal{L}_R=\frac{1}{2}(\partial_{\mu}\phi)^2-\frac{1}{2}m^2\phi^2-\frac{1}{6}g\mu^{3-d/2}\phi^3+ j\mu^{d/2-3}\phi,
\label{reno_lag}
\end{equation}
\begin{equation}
\mathcal{L}_{CT}=\frac{1}{2}{\delta}z_1(\partial_{\mu}\phi)^2-\frac{1}{2}{\delta}m^2\phi^2-\frac{1}{6}{\delta}g\mu^{3-d/2}\phi^3+\delta j \mu^{d/2-3}\phi,
\label{CTLag}
\end{equation}
where we have included the arbitrary parameter $\mu$ with mass dimension at two places so that both \(m^4/j\) and \(g\) dimensionless, given the
dimensions,
\(dim(\phi_0)=d/2-1,\ dim(m_0^2)=2,\ dim(g_0)=3-d/2,\ dim(j_0)=d/2+1\).

In the minimal subtraction scheme, the counterterms are series of poles that diverge when $d=6$\cite{t_hooft_dimensional_1973}. 
\begin{align}
g_0\mu^{d/2-3}&=g+\sum_{\nu=1}^{\infty}\frac{a_{\nu}(g,m,\mu)}{(d-6)^{\nu}},\label{bareLaurent1}\\
m_0&=m+m\sum_{\nu=1}^{\infty}\frac{b_{\nu}(g,m,\mu)}{(d-6)^{\nu}},\label{bareLaurent2}\\
z_1&=1+\sum_{\nu=1}^{\infty}\frac{c_{\nu}(g,m,\mu)}{(d-6)^{\nu}},\label{bareLaurent3}\\
j_0\mu^{3-d/2}&=j+m^4\sum_{\nu=1}^{\infty}\frac{e_\nu(g,m,\mu,m^4/j)}{(d-6)^\nu},\label{bareLaurent4}
\end{align}
where \(z_1\) is given by
\begin{equation}
\phi_0^2\equiv z_1\phi^2=(1+\delta z_1)\phi^2.
\end{equation}
The terms in the residues $a_\nu$, $b_\nu$, $c_\nu$, $e_\nu$ turn out to be independent of $m$, $j$ or $\mu$ because (i) they must be dimensionless; therefore $m$, $\mu$ must occur only as ratios ($j$ doesn't occur in any of the counterterms because the vertex $j\phi$ cannot occur in any loop)\cite{collins_new_1974}, and (ii) they can only include $m^2$ in its polynomials and $\mu$ in its logarithms\cite{t_hooft_regularization_1972}. 

The renormalisation group functions for the parameters are defined so:
\begin{align}
	\flow{g} &=(d/2-3)g+\beta(g),\\
	\flow {m^2} &= m^2\delta(g),\\
	\flow {z_1} &= 2\gamma(g) z_1,\\
	\flow j &= -(d/2-3)j +\eta(g,m^4/j).
\end{align}
The dependence of $\eta$ on $m^4/j$ can be removed by a slight redefinition, but it is not very convenient for our purpose (cf. \eqref{bar_eta}):
\begin{equation}
	\flow j = -(d/2-3)j + j\gamma(g) + m^4\bar{\eta}(g).
\end{equation}

To one-loop order these counterterms and RG functions are calculated in appendix \ref{appendix:oneloop}.


\section{Renormalisation of composite operators}
\label{compositeOperators}


We express the bare parameters in terms of the renormalised parameters in exponential form for convenience\cite{brown_dimensional_1980}. While we express the infinite series of divergences in a compact form, we nowhere make use of this \emph{sum of divergences} in the sense of \cite{t_hooft_dimensional_1973}.  

\begin{itemize}
\item
  For \(g\),
  \begin{equation}
  g_0=\mu^{3-d/2}g\exp\{U(g;d)\},
  \label{g_0-g}
  \end{equation}
  where on direct integration after differentiating with \(\mu\),
  \begin{equation}
  U(g;d)=-\int_0^g\frac{\mathrm dg^\prime}{g^\prime}\frac{\beta(g^\prime)}{g^\prime(d/2-3)+\beta(g^\prime)}.
  \end{equation}
\item
  For \(m^2\),
\begin{equation}
m_0^2\equiv m^2/z_2=m^2\exp\{V(g;d)\},
\label{m_0-m}
\end{equation}
where
\begin{equation}
V(g;d)=-\int_0^g\mathrm dg^\prime\frac{\delta(g^\prime)}{g^\prime(d/2-3)+\beta(g^\prime)}.
\end{equation}
\item
  For \(z_1\),
  \begin{equation}
  z_1=\exp\{W(g;d)\}, \label{z_1}
  \end{equation}
  where
  \begin{equation}
  W(g;d)=\int_0^g\mathrm dg^\prime\frac{2\gamma(g^\prime)}{g^\prime(d/2-3)+\beta(g^\prime)}.
  \end{equation}
\item
  For \(j\): Writing the counterterms in exponential form is not possible for the linear parameter $j$, because the counterterms are proportianal to $m^4$ and do not depend on $j$. We write the counterterms in the following form.
  \begin{equation}
  j_0\sqrt{z_1}=\big(j+\frac{m^4}gX(g;d)\big)\mu^{d/2-3},
  \label{j_0-j}
  \end{equation}
  Here $X(g;d)$ is an ascending series of poles (cf. \eqref{bareLaurent4}). We wish to obtain an integral expression for it similar to $U$, $V$, $W$ above.
  
  Now $j_0$ is RG invariant,
  \begin{align*}
  0 &= \mu\frac{\mathrm d}{\mathrm d\mu}\bigg[\frac1{\sqrt{z_1}}(j+\frac{m^4}gX)\bigg]\mu^{d/2-3} + (d/2-3)j_0\\
  0 &= -\frac1{\sqrt{z_1}}\gamma\times (j+\frac{m^4}gX) + \frac1{\sqrt{z_1}}\bigg[\mu\frac{\mathrm dj}{\mathrm d\mu} + 2\frac{m^4}g\delta(g) X + \frac{m^4}g\bigg(\frac{\partial X}{\partial g}-\frac1gX\bigg)\mu\frac{\mathrm dg}{\mathrm d\mu}\bigg] \\
  &\quad+ \frac1{\sqrt{z_1}}(d/2-3)(j+\frac{m^4}gX)\\
  0&=-\gamma (j+\frac{m^4}gX) + j\big(-(d/2-3) + \eta\big) + 2\frac{m^4}gX\delta(g) + \frac{m^4}g\frac{\partial X}{\partial g}\mu\frac{\mathrm dg}{\mathrm d\mu}\\
  &\quad-\frac{m^4}{g^2}X\big[g(d/2-3)+\beta] + (d/2-3)(j+\frac{m^4}gX)\\
  0&=\frac{m^4}g\frac{\partial X}{\partial g}\mu\frac{\mathrm dg}{\mathrm d\mu} + \big(2\delta(g)-\gamma-\beta/g\big)\frac{m^4}gX + j(\eta-\gamma).
  \end{align*}
  We use an integrating factor to simplify this differential equation
  (recalling \(\mu\frac{\partial g}{\partial \mu}=g(d/2-3)+\beta\)):
\begin{align}
  X\to \bar{X} &= X\exp\{Y(g;d)\},\\
  \text{where } Y(g;d) &\equiv \int_0^g \frac{2\delta(g^\prime)-\gamma(g^\prime)-\beta(g^\prime)/g^\prime}{g^\prime(d/2-3)+\beta(g^\prime)}\ \mathrm dg^\prime.\nonumber
  \end{align}
  The differential equation for \(\bar{X}\) is then,
  \begin{align}
	\gder{\bar{X}}=\frac{jg}{m^4}\times\frac{\gamma-\eta}{g(d/2-3)+\beta} \times \exp\{Y(g;d)\}\nonumber\\
	  \implies \bar{X}=\int_0^g\frac{jg^\prime}{m^4}\times\frac{\gamma(g^\prime)-\eta(g^\prime)}{g^\prime(d/2-3)+\beta(g^\prime)} \times \exp\{Y(g^\prime;d)\}\ \mathrm{d}g^\prime,
	\end{align}
	and,
	\begin{equation}
	\boxed{\therefore X=\exp\{-Y(g;d)\}\int_0^g\frac{jg^\prime}{m^4}\times\frac{\gamma(g^\prime)-\eta(g^\prime)}{g^\prime(d/2-3)+\beta(g^\prime)} \exp\{Y(g^\prime;d)\}\ \mathrm{d}g^\prime}. \label{j-poles}
	\end{equation}
\end{itemize}

A couple of comments:
\begin{itemize}
\item
  \(X\) is supposed to be a function of \(g\) and \(d\) alone. This
  means \(\gamma(g)-\eta(g,m^4/j)\equiv -(m^4/j)\bar{\eta}(g)\), for
  some new function \(\bar{\eta}(g)\). 
  \begin{equation}
  \mu \frac{\mathrm dj}{\mathrm d\mu} = -j(d/2-3) + j\gamma(g) + m^4\bar{\eta}(g).
  \label{bar_eta}
  \end{equation}
\item
  The value of \(X\) can be verified by comparing to \(\delta j\) calculated in the appendix
  \eqref{deltaj}. To the smallest order in \((d-6)^{-1}\), we expand \(X\).

  \begin{align*}
  X &= \{1-Y+\ldots\}\int_0^g \frac{jg^\prime}{m^4}\times\frac{\gamma(g^\prime)-\eta(g^\prime)}{g^\prime(d/2-3)}\bigg\{1-O\bigg(\frac{\beta(g^\prime)}{g^\prime(d/2-3)}\bigg)\bigg\}\{1+Y\}\ \mathrm dg^\prime\\
  &= \int_0^g \frac{j}{m^4}\times\frac{\gamma(g^\prime)-\eta(g^\prime)}{(d/2-3)}\ \mathrm dg^\prime \quad+ higher\ poles\\
  &= \frac{j}{m^4(d/2-3)}\int_0^g \bigg\{\frac{g^{\prime2}}{12(4\pi)^3} - \frac{g^\prime}{2(4\pi)^3}\bigg(\frac {g^\prime}6+\frac{m^4}{j}\bigg)\bigg\}\ \mathrm dg^\prime \quad+higher\ poles\\
  &= -\frac{1}{(d/2-3)}\int_0^g \frac{g^\prime}{2(4\pi)^3}\ \mathrm dg^\prime \quad+higher\ poles\\
  &= -\frac{g^2}{2(4\pi)^3(d-6)} \quad+higher\ poles
  \end{align*}
  where we have substituted for the values of \(\gamma\) and \(\eta\) to
  lowest order. This matches the value for $\delta j$.
\end{itemize}

\subsection{Relations between derivatives}
We will need the relations between derivatives w.r.t. the bare and renormalised parameters.

\begin{itemize}
\item
  We have the straightforward relation for \(\mathrm dg_0-\mathrm dg\)
  from \eqref{g_0-g}.
  \begin{equation}
  \mathrm dg_0=g_0\frac{d/2-3}{g(d/2-3)+\beta}\mathrm dg.
  \end{equation}
\item
  For \(m^2\), from \eqref{m_0-m},
  \begin{equation}
  \mathrm dm_0^2=m_0^2\bigg[\frac{\mathrm dm^2}{m^2}-\frac{\delta \mathrm dg}{g(d/2-3)+\beta}\bigg].
  \end{equation}
\item
  For \(j\), then, from \eqref{j_0-j},
  \begin{align}
  \sqrt{z_1}\mathrm dj_0 + \frac12 j_0\sqrt{z_1}\frac{\partial W}{\partial g}\mathrm dg= &\bigg(dj + 2\frac{m^2X}{g}\mathrm dm^2 + \frac{m^4}{g}\Big(\frac{\partial X}{\partial g}-\frac{X}{g}\Big)\mathrm dg\bigg)\mu^{d/2-3}\nonumber\\
  \mathrm dj_0= &\frac{\mu^{d/2-3}}{\sqrt{z_1}}\mathrm dj + \frac{2m^2X\mu^{d/2-3}}{g\sqrt{z_1}}\mathrm dm^2\nonumber \\
  &+ \bigg[\frac{m^4\mu^{d/2-3}}{g\sqrt{z_1}}\bigg(\frac{\partial X}{\partial g}-\frac{X}{g}\bigg)-\frac12 j_0\frac{\partial W}{\partial g}\bigg]\mathrm dg
  \end{align}
\end{itemize}

From the above relations, we have, for any function $F$ of these parameters,
\begin{align}
\frac{\partial F}{\partial j} = &\frac{\mu^{d/2-3}}{\sqrt{z_1}}\frac{\partial F}{\partial j_0},\label{j-deriv}\\
\frac{\partial F}{\partial m^2}=&\frac{m_0^2}{m^2}\frac{\partial F}{\partial m_0^2} + \frac{2m^2X\mu^{d/2-3}}{g\sqrt{z_1}}\frac{\partial F}{\partial j_0},\label{m^2-deriv}\\
\frac{\partial F}{\partial g}=&\frac1{g(d/2-3)+\beta}\bigg[(d/2-3)g_0\frac{\partial F}{\partial g_0}-m_0^2\delta(g)\frac{\partial F}{\partial m_0^2}\bigg]\nonumber\\
&+ \bigg[\frac{m^4\mu^{d/2-3}}{g\sqrt{z_1}}\bigg(\frac{\partial X}{\partial g}-\frac{X}{g}\bigg)-\frac12j_0\frac{\partial W}{\partial g}\bigg]\frac{\partial F}{\partial j_0}\label{g-deriv}.
\end{align}
By applying these rules on correlation functions, we will find relations between bare and
renormalised composite operators. \eqref{j-deriv} just gives the
already known wavefunction renormalisation relation, \(\phi_0=\sqrt{z_1}\phi\).

\subsection{\(\phi^2\) operator renormalisation}
\label{phi2}
A composite operator can mix with operators of same or dimension. So
\(\phi_0^2\) mixes with \(\phi_0\) and \(\partial^2\phi_0\).
Arranging the bare operators in a column matrix,
\begin{equation}
Q_0(x)=
\begin{pmatrix}
\frac12m_0^2\phi_0^2\\j_0\phi_0\\\partial^2\phi_0
\end{pmatrix},
\end{equation}
we can cast the renormalisation equation into matrix form,
\begin{equation}
Q_0(x)=Z\times[Q](x),
\label{phi2vectoreq}
\end{equation}
where \([Q](x)\) is the column matrix of the renormalised operators,
\begin{equation}
[Q](x)=
\begin{pmatrix}
\frac12m^2[\phi^2]\\\mu^{d/2-3}j\phi\\\partial^2\phi
\end{pmatrix},
\end{equation}
and \(Z\) is a \(3\times3\) matrix,
\begin{equation}
Z=
\begin{pmatrix}
A&B&C\\0&1+\frac{m^4}{jg}X&0\\0&0&\sqrt{z_1}
\end{pmatrix}
\label{phi2matrix},
\end{equation}
where $B(g,m^2,j,\mu;d)$ and $C(g,m^2,j,\mu;d)$ are series of poles in $(d/2-3)$, and $A(g;d)$ is $1+$ series of poles. ($[\phi^2]$ can be normalised so that the finite part of $A$ is 1.)

Consider a renormalised correlation function \(G_N\) given by
\begin{equation}
G_N(x_1,\dots,x_N)=i^{N/2}z_1(g)^{-N/2}\int\mathcal{D}\phi_0\ \phi_0(x_1)\dots\phi_0(x_N)\exp\{i\int\mathrm d^dx\ \mathcal{L}\},
\end{equation}
with the Lagrangian \(\mathcal{L}\) given by \eqref{lagrangian}.
From \eqref{m^2-deriv}, (\(G_N(O)\) is the correlation function with
an operator \(O\) inserted),
\begin{align}
m^2\frac{\partial G_N}{\partial m^2} &= i\int\mathrm d^dx\big\{-G_N(\frac12m_0^2\phi_0^2(x))+2\frac{m^4}{g\sqrt{z_1}}X\mu^{d/2-3}G_{N}(\phi_0(x))\big\},\nonumber\\
\text{from }\eqref{phi2vectoreq} \text{ and }\eqref{phi2matrix},\nonumber\\
&= -i\int\mathrm d^dx\big\{AG_N(\frac12m^2[\phi^2(x)])+(Bj-2\frac{m^4}{g}X)\mu^{d/2-3}G_{N}(\phi(x))\big\}
\end{align}
where we have expressed the bare operators in terms of the
renormalised ones using \(Z\). (The total derivative operator vanishes
under integration. So we can't use this to find \(C\).)

Now LHS of above equation is finite, so RHS must be too. Since the operators on the right are linearly independent, \(A\) and
\(Bj-2\frac{m^4}{g}X\) must be finite. \(X \) is clearly not finite;
\(m^4,\ j\) are. So, \(B=-2\frac{m^4}{jg}X\) since it has no finite part (minimal subtraction).
Thus,
\begin{equation}
A=1;\ B=2\frac{m^4}{jg}X=\frac{2\delta j}j.
\label{phi2vectoreqresult}
\end{equation}

$C$ can be given an integral expression in terms of a new anomalous dimension. We first define the anomalous dimesion matrix:
\begin{align}
\mu \frac{\mathrm d}{\mathrm d\mu}Z &\equiv Z\Gamma\nonumber\\
\Gamma&=
\begin{pmatrix}
0&2(\gamma-\eta)&?\\
0&\gamma-\eta&0\\
0&0&\gamma
\end{pmatrix}
\label{Gamma}
\end{align}
\(\Gamma\) must be finite; this can be seen from differentiating \eqref{phi2vectoreq}.
\begin{align*}
	\flow{Z}[Q](x)+Z\flow{[Q](x)}=0,\\
	\implies Z^{-1} \flow{Z}[Q](x)+\flow{[Q](x)}=0,\\
	\therefore \flow{[Q](x)}=-\Gamma[Q](x).
\end{align*}
$\flow{[Q](x)}$ and $[Q](x)$ are finite, therefore so is $\Gamma$.

Defining the missing element to be \(m^2\mu^{d/2-3} \zeta\), the
equation to be satisfied by \(C\) and \(\zeta\) is
\begin{equation}
\mu\frac{\mathrm dC}{\mathrm d\mu} = m^2\mu^{d/2-3}\zeta+C\gamma.
\label{Cflow}
\end{equation}
Dimensionally, \(C\) should have the form,
\begin{equation}
C=m^2\mu^{d/2-3}\bar{C}/g,
\end{equation}
with \(\bar{C}\) only a function of \(g\) and \(d\).
\begin{equation}
\therefore g(d/2-3 + \beta/g)\frac{\partial \bar{C}}{\partial g} + (\delta-\gamma-\beta/g)\bar{C} = g\zeta.
\end{equation}
Thus, \(\zeta\) is also a function of \(g\) alone, (since it is finite).
\begin{gather*}
\bar{C} =\exp\{-D(g;d)\}\int_0^g \mathrm d g^\prime\frac{\zeta(g^\prime)}{d/2-3+\beta(g^\prime)/g^\prime} \exp\{D(g^\prime;d)\},\\
\textrm{where }D(g;d)\equiv \int_0^g \mathrm d g^\prime\frac{\delta(g^\prime)-\gamma(g^\prime)-\beta(g^\prime)/g^\prime}{d/2-3+\beta(g^\prime)/g^\prime}.
\end{gather*}
And thus,
\begin{equation}
C=\frac{m^2\mu^{d/2-3}}{g}\exp\{-D(g;d)\}\int_0^g \mathrm d g^\prime\frac{\zeta(g^\prime)}{d/2-3+\beta(g^\prime)/g^\prime} \exp\{D(g^\prime;d)\}.
\label{C}
\end{equation}
This \(\zeta\) is an entirely new anomalous dimension, associated with
\(\partial^2\phi\).

\subsection{\(\phi^3\) operator renormalisation}
\label{phi3}
The dimension 6 and lower operators are
\(\frac16g\phi^3,\ \frac12m^2\phi^2,\ \partial^4 \phi,\ \phi\partial^2\phi,\ \partial^2\phi^2,\ j\phi, \partial^2\phi\).

It is convenient to work in terms of the operator related to field equation, labeled \(E_0(x)\), defined by
\begin{equation}
E_0(x)\equiv-\phi_0\frac{\delta S}{\delta\phi_0}=\phi_0\{(\partial^2+m_0^2)\phi_0+\frac12g_0\phi_0^2-j_0\}.
\label{eomop}
\end{equation}
Insertion of this operator in a correlation function just gives a
multiplicative factor\cite{brown_1992}:
\begin{equation}
G_N(x_1,\dots,x_N;iE_0(x))=\sum_{\alpha=1}^N \delta(x-x_\alpha)G_N(x_1,\dots,x_N). \label{S-D}
\end{equation}
Thus, \(E_0(x)\) is finite, and does not renormalise:
\begin{equation}
E_0(x)=[E](x),
\end{equation}
and we can work with it instead of \(\phi_0\partial^2\phi_0\).

Defining $Q_0$, $[Q]$, and $Z$ similar to before; (we use the same symbols, but hope this doesn't cause confusion; the earlier calculation only feeds into this one, and isn't used later).
\begin{equation}
Q_0(x)\equiv\begin{pmatrix}
\frac16g_0\phi_0^3\\ \frac12m_0^2\phi_0^2\\ j_0\phi_0\\ \partial^2\phi_0\\ E_0(x)\\ \partial^4 \phi_0\\ \partial^2\phi_0^2
\end{pmatrix};
\quad [Q](x)\equiv\begin{pmatrix}
\frac16g\mu^{3-d/2}[\phi^3]\\ \frac12m^2[\phi^2]\\ j\mu^{d/2-3}\phi\\ \partial^2\phi\\ [E](x)\\ \partial^4 \phi\\ \partial^2[\phi^2]
\end{pmatrix},
\end{equation}
with
\begin{equation}
Q_0(x)=Z[Q](x).\label{3-matrix-eq}
\end{equation}
The matrix \(Z\) must have the structure, filling from earlier values,
\begin{equation}
Z=\begin{pmatrix}
1+a_1 & a_2 & a_3 & a_4 & a_5 & a_6 & a_7\\
0 & 1 & B & C & 0 & 0 & 0\\
0 & 0 & 1+\frac{m^4}{jg}X & 0 & 0 & 0 & 0\\
0 & 0 & 0 & \sqrt{z_1} & 0 & 0 & 0\\
0 & 0 & 0 & 0 & 1 & 0 & 0\\
0 & 0 & 0 & 0 & 0 & \sqrt{z_1} & 0\\
0 & 0 & 0 & \frac{2j\mu^{d/2-3}}{m_0^2}B & 0 & \frac2{m_0^2}C & \frac{m^2}{m_0^2}
\end{pmatrix},
\end{equation}
where \(a_i\) have no finite components.

From \eqref{g-deriv} we have,
\begin{align}
\frac{\partial G_N}{\partial g}=&\frac{z_1^{-N/2}}{g(d/2-3)+\beta}\bigg[-N\gamma+(d/2-3)g_0\frac\partial{\partial g_0}-m_0^2\delta\frac\partial{\partial m_0^2}\bigg]G_N^{(0)}\nonumber\\
&+\bigg[\frac{m^4\mu^{d/2-3}}{g\sqrt{z_1}}\bigg(\frac{\partial X}{\partial g}-\frac{X}{g}\bigg)-\frac12j_0\frac{\partial W}{\partial g}\bigg]\frac{\partial G_N}{\partial j_0},
\end{align}
where we have written \(G_N=G_N^{(0)}z_1^{-N/2}\) to obtain the
\(-N\gamma\) term,
\begin{equation}
\because \frac{\partial}{\partial g}z_1^{-N/2}=-\frac N2 z_1^{-N/2+1}\frac{\partial z_1}{\partial g}=-\frac N2 z_1^{-N/2}\frac{2\gamma}{g(d/2-3)+\beta} \textrm{ from } \eqref{z_1}.\\
\end{equation}
From \eqref{j-poles},
\begin{equation}
\frac{\partial X}{\partial g}=-\frac{2\delta-\gamma-\beta/g}{g(d/2-3)+\beta}X+\frac{jg}{m^4}\times\frac{\gamma-\eta}{g(d/2-3)+\beta},
\end{equation}
and
\begin{equation}
\frac{\partial W}{\partial g}=\frac{2\gamma}{g(d/2-3)+\beta}.
\end{equation}
The derivatives w.r.t. \(g_0,\ m_0^2,\) and \(j_0\) effect insertions of
\(-\frac16\phi_0^3,\ -\frac12\phi_0^2\), and \(\phi_0\) respectively.

Further, using \eqref{S-D}, we can write the \(-NG_N\) as an insertion
of \(E_0\). Thus,
\begin{align*}
\frac{\partial G_N}{\partial g}&=\frac{i}{g(d/2-3)+\beta}\int\mathrm d^dx\bigg[-\gamma G_N(E_0(x))-(d/2-3)G_N(\frac16g_0\phi_0^3(x))\\
&\quad +\delta G_N(\frac12m_0^2\phi_0^2(x)) +\bigg(\frac{m^4\mu^{d/2-3}}{g\sqrt{z_1}} (\gamma+\beta/g-2\delta)X+\frac{j\mu^{d/2-3}}{\sqrt{z_1}}(\gamma-\eta)\\
&\qquad-\frac{m^4\mu^{d/2-3}}{g\sqrt{z_1}}((d/2-3)+\beta/g)X-\gamma j_0 \bigg) G_N(\phi_0(x))\bigg]\\
&=\frac{i}{g(d/2-3)+\beta}\int\mathrm d^dx\bigg[-\gamma G_N(E_0(x))-(d/2-3)G_N(\frac16g_0\phi_0^3(x))\\
&\quad +\delta G_N(\frac12m_0^2\phi_0^2(x)) -\bigg(\frac{m^4\mu^{d/2-3}}{g\sqrt{z_1}} (2\delta+d/2-3)X+\frac{1}{\sqrt{z_1}}j\mu^{d/2-3}\eta\bigg) G_N(\phi_0(x))\bigg].
\end{align*}
Now we can use \eqref{3-matrix-eq} to replace bare operators
with renormalised ones. The ones with derivatives vanish under integral.
\begin{align*}
\frac{\partial G_N}{\partial g}&=\frac{i}{g(d/2-3)+\beta}\int\mathrm d^dx\bigg[-\gamma G_N([E](x))\\
&\quad -(d/2-3)\{(1+a_1)G_N(\frac16g[\phi^3](x)) + a_2 G_N(\frac12m^2[\phi^2](x))\\
&\qquad +a_3G_N(j\mu^{d/2-3}\phi(x))+a_5G_N([E](x))\}\\
&\quad +\delta \{G_N(\frac12m^2[\phi^2](x))+\frac{2m^4X}{jg}G_N(j\mu^{d/2-3}\phi(x))\} \\
&\quad -\bigg(\frac{m^4\mu^{d/2-3}}{g} (2\delta+d/2-3)X+j\mu^{d/2-3}\eta\bigg) G_N(\phi(x))\bigg]\\
&=\frac{i}{g(d/2-3)+\beta}\int\mathrm d^dx\bigg[-(d/2-3)(1+a_1)G_N(\frac16g[\phi^3](x))\\
&\quad - ((d/2-3)a_2 - \delta) G_N(\frac12m^2[\phi^2](x))- ((d/2-3)a_5+\gamma) G_N([E](x))\\
&\quad -((d/2-3)(a_3+\frac{m^4X}{jg})+\eta)G_N(j\mu^{d/2-3}\phi(x))\bigg]
\end{align*}
Now the LHS is clearly finite, and so are the correlation functions on the RHS
with insertions of renormalised operators. So, the coefficients of each
of these correlation functions, (which are linearly independent), must be
finite. But the factor outside the integral is an infinite series of divergences. Therefore, coefficient of each of the correlation functions inside the integral must either vanish or cancel the factor outside exactly.
Thus, we have
\begin{equation}
	\begin{aligned}
	a_1&=\frac{\beta/g}{d/2-3};\\
	a_2&=\frac{\delta}{d/2-3};\\
	a_3&=-\frac{\eta}{d/2-3}-\frac{m^4X}{jg}=-\frac{j\gamma+m^4\bar{\eta}}{j(d/2-3)}-\frac {m^4X}{jg};\\
	a_5&=-\frac{\gamma}{d/2-3}.
	\end{aligned}
	\label{matrixrow1}
\end{equation}

For the rest, we need to define the anomalous dimension matrix
\(\Gamma\).
\begin{equation}
\mu\frac{\mathrm d }{\mathrm d\mu}Z\equiv Z\Gamma.
\end{equation}
Then we have
\begin{equation}
\Gamma=
\begin{pmatrix}
g\gder{}(\beta/g) & g\gder\delta & -g\gder\eta & ? & -g\gder \gamma & ? & ?\\
0 & 0 & 2(\gamma-\eta) & m^2\zeta\mud & 0 & 0 & 0\\
0 & 0 & \gamma-\eta  & 0 & 0 & 0 & 0\\
0 & 0 & 0 & \gamma & 0 & 0 & 0\\
0 & 0 & 0 & 0 & 0 & 0 & 0\\
0 & 0 & 0 & 0 & 0 & \gamma & 0\\
0 & 0 & 0 & \frac{4j\mud}{m^2}(\gamma-\eta) & 0 & 2\zeta\mud & \delta
\end{pmatrix}
\end{equation}
Defining the three missing elements \(\zeta_4m^2\mud,\ \zeta_6\mud\), and \(\zeta_7\), respectively, so that we get
\begin{align}
	\flow {a_4}&=(1+a_1)\zeta_4m^2\mud+m^2a_2\mud\zeta+a_4\gamma+4a_7\frac j {m^2}\mud(\gamma-\eta),\label{a4flow}\\
	\flow {a_6} &= (1+a_1)\zeta_6\mud + a_6\gamma + 2\mud a_7\zeta,\label{a6flow}\\
	\flow {a_7} &= (1+a_1)\zeta_7 + a_7 \delta.\label{a7flow}
\end{align}
These elements $\zeta_4$, $\zeta_6$, and $\zeta_7$ are again new anomalous dimensions similar to $\zeta$, associated with the operators $\partial^2 \phi$, $\partial^4 \phi$, and $\partial^2 [\phi^2]$ respectively. They turn out to be zero to $O(g^2)$ from one-loop calculation in appendix \ref{oneloopreno}.

\(a_7\) is simple:
\begin{equation}
a_7= z_2(g;d)\int_0^g\frac{\dd g\pr}{g\pr}z_2(g\pr;d)^{-1} \frac{\zeta_7(g\pr)}{d/2-3},
\label{a_7}
\end{equation}

where \(z_2\equiv m^2/m_0^2\) is given by \eqref{m_0-m}.
\(a_6\) can be written in terms of \(a_7\), and then \(a_4\) in terms of
\(a_6\) and \(a_7\), (but they look messy and we won't be needing them). This calculation is verified against one-loop Feynman diagram computation of counterterms for $\phi^3$ in the appendix \ref{oneloopreno}.

\section{Tensor Operators}
\label{tensorOperators}

Having renormalised the scalar composite operators, we can now do this to tensors. Consider traceless tensors since renormalisation doesn't mix them with
scalar operators multiplied by the metric tensor. We define the differential operator
\begin{equation}
t_{\mu\nu}\equiv d \partial_\mu\partial_\nu-\gmn \partial^2.
\end{equation}
We work with the bare operator \(\phi_0\tmn\phi_0\) which has the
renormalised composite operator expansion
\begin{equation}
\phi_0\tmn\phi_0(x)=A_1[\phi\tmn\phi](x)+A_7\tmn[\phi^2](x)+A_4 \tmn\phi(x)+A_6\tmn\partial^2\phi(x),
\label{tensorrenorm}
\end{equation}
where \(A_1\) must be unity plus poles in \((d/2-3)\), and
\(A_4, A_6, A_7\) must be poles in \((d/2-3)\) alone. There are no
further operators possible with dimension lower than 6 and two tensor
indices.

These coefficients can be evaluated by examining
\begin{equation}
\partial^\nu (\phi_0\tmn\phi_0)=A_1\partial^\nu[\phi\tmn\phi]+A_7(d-1)\partial_\mu\partial^2[\phi^2]+A_4 (d-1)\partial_\mu\partial^2\phi+A_6(d-1)\partial_\mu\partial^4\phi.
\label{exam}
\end{equation}
The L.H.S. can be expressed in terms of known quantities as follows.
\begin{align}
\partial^\nu (\phi_0\tmn\phi_0)&=d\partial^\nu (\phi_0\partial_\mu\partial_\nu\phi_0)-\partial_\mu(\phi_0\partial^2\phi_0)\nonumber\\
&\equiv a\partial_\mu \partial^2 \phi_0^2 + b \partial_\mu m_0^2\phi_0^2 + c \partial_\mu \frac16 g_0\phi_0^3 + e \partial_\mu E_0 + f E_{0\mu} + h \partial_\mu j_0\phi_0,
\label{expansion}
\end{align}
where \(E_{0\mu}\) is similar to the equation of motion operator
\(E_0\), and defined by
\begin{equation}
E_{0\mu}\equiv(\partial_\mu\phi_0)\{(\partial^2+m_0^2)\phi_0+\frac12g_0\phi_0^3-j_0\}.
\end{equation}
Similar to \(E_0\), it obeys the property
\begin{equation}
G_N(x_1,\ldots,x_N;iE_{0\mu}(x))=\sum_{a=1}^N\delta(x-x_a)\frac\partial{\partial x_a^\mu}G_N(x_1,\ldots,x_N).
\end{equation}
Thus, it is finite too:
\begin{equation}
E_{0\mu}(x)=[E_\mu](x).
\end{equation}

We must have, (recall \(E_0\) from \eqref{eomop}),
\begin{align*}
\frac c {3!}+\frac e2+\frac f{3!}&=0,\\
b+e+\frac f2&=0,\\
h-e-f&=0.
\end{align*}
Expressing the LHS of \eqref{expansion} in terms of operators that
could be part of RHS,
\begin{equation}
\partial^\nu (\phi_0\tmn\phi_0)=(d/2-1)\partial_\mu(\phi_0\partial^2\phi_0)+\frac d4 \partial_\mu \partial^2\phi_0^2-d(\partial_\mu\phi_0)\partial^2\phi_0,\nonumber
\end{equation}
\begin{equation}
\therefore e=d/2-1;\quad f=-d;\quad a=\frac d4;\nonumber\\
\& \therefore c=-(d/2-3);\quad b=1;\quad h=-d/2-1.
\end{equation}
Thus,
\begin{align}
\partial^\nu (\phi_0\tmn\phi_0)=&\frac d4 \partial_\mu \partial^2 \phi_0^2 + \partial_\mu m_0^2\phi_0^2 -(d/2-3) \partial_\mu \frac16 g_0\phi_0^3 + (d/2-1) \partial_\mu E_0 -d E_{0\mu}\nonumber\\
& -(d/2+1) \partial_\mu j_0\phi_0,
\label{expansion2}
\end{align}

Now consider \eqref{exam}. \(E_{0\mu}\) is finite and doesn't mix with
any of the other operators in above equations.
\(\partial^\nu (\phi_0\tmn\phi_0)\) has \(E_{0\mu}\) and
\(\partial^\nu[\phi\tmn\phi]\) has \([E_\mu]\), but since they are both
equal, and rest of the terms in \eqref{exam} don't contain
\([E_\mu]\), \(A_1\) must be 1.

Now we know the renormalisation of terms in \eqref{expansion2}, and
thus we have, using \eqref{3-matrix-eq},
\begin{align}
\partial^\nu[\phi\tmn\phi]&=-\{(d/2-3)+\beta/g\}\frac16g\mu^{3-d/2}\partial_\mu[\phi^3]+(1-\delta/2)m^2\partial_\mu[\phi^2]+\{d/2-1+\gamma\}\partial_\mu[E]
\label{tensordivergence}\\
&\quad -d[E_\mu] + (\eta-d/2-1)j\mu^{d/2-3}\partial_\mu\phi+ (d/4+\xi_7)\partial_\mu\partial^2[\phi^2]+\xi_4\partial_\mu\partial^2\phi+\xi_6\partial_\mu\partial^4\phi,\nonumber
\end{align}
where \(\xi_4,\xi_6,\xi_7\) are finite and are determined by
\begin{align}
\frac d4+\xi_7+(d-1)A_7=\frac d4z_2-a_7(d/2-3);\\
\xi_4+(d-1)A_4=d\times m^2z_2X\mu^{d/2-3}/g+2C-(d/2-3)a_4;\\
\xi_6+(d-1)A_6=\frac d4\times\frac{2C}{m_0^2}-(d/2-3)a_6.
\end{align}
We have completely determined \eqref{tensorrenorm} in terms of the earlier defined anomalous dimensions $\zeta$, $\zeta_4$, $\zeta_6$, $\zeta_7$ and the counterterms $a_4$, $a_6$, $a_7$, and $C$. (Refer \eqref{a4flow}, \eqref{a6flow}, \eqref{a7flow}, \eqref{C}.)

\section{Energy Momentum Tensor}
\label{emtensor}
The canonical energy momentum tensor is given by
\begin{equation}
T^C_{\mu\nu}\equiv\frac{\partial \mathcal L}{\partial(\partial^\mu\phi_0)}\partial_\nu\phi_0-\gmn\mathcal L.
\end{equation}
\begin{equation}
\text{i.e., }T^C_{\mu\nu}=\partial_\mu\phi_0\partial_\nu\phi_0-\gmn\Big\{\frac12(\partial\phi_0)^2-\frac12m_0^2\phi_0^2-\frac1{3!}g_0\phi_0^3+j_0\phi_0\Big\}.
\end{equation}

This is not finite. We can make it finite by adding terms involving
\(\smn\equiv\partial_\mu\partial_\nu-\gmn\partial^2\), since it is
structurally conserved. \((\partial^\mu \smn O = 0.)\) We have
\begin{equation}
\partial_\mu\phi_0\partial_\nu\phi_0=\partial_\mu\partial_\nu\frac12\phi_0^2-\phi_0\partial_\mu\partial_\nu\phi_0,
\end{equation}
\begin{equation}
\therefore T^C_{\mu\nu}=\partial_\mu\partial_\nu\frac12\phi_0^2-\phi_0\partial_\mu\partial_\nu\phi_0-\gmn\Big\{\frac14\partial^2\phi_0^2-\frac12\phi_0\partial^2\phi_0-\frac12m_0^2\phi_0^2-\frac1{3!}g_0\phi_0^3+j_0\phi_0\Big\}.
\end{equation}

We express $\partial_\mu\partial_\nu\phi_0^2$ and $\gmn\partial^2\phi_0^2$ in terms of \(\tmn\) and \(\smn\) using
\begin{equation}
a\partial_\mu\partial_\nu-b\gmn \partial^2=\frac{a-b}{d-1}\tmn+\frac{db-a}{d-1}\smn,
\end{equation}
and the operators $\phi_0\partial_\mu\partial_\nu\phi_0$ and $\phi_0\partial^2\phi_0$ in terms of $\phi_0\tmn\phi_0$ and $E_0$.
\begin{align*}
T^C_{\mu\nu}&=\frac1{4(d-1)}\tmn\phi_0^2+\frac{d-2}{4(d-1)}\smn\phi_0^2-\frac1d\phi_0\tmn\phi_0\\
&\quad-\gmn\Big\{\Big(\frac1d-\frac12\Big)\phi_0\partial^2\phi_0-\frac12m_0^2\phi_0^2-\frac1{3!}g_0\phi_0^3+j_0\phi_0\Big\}\\
&=\frac1{4(d-1)}\tmn\phi_0^2+\frac{d-2}{4(d-1)}\smn\phi_0^2-\frac1d\phi_0\tmn\phi_0\\
&\quad-\gmn\Big\{\Big(\frac1d-\frac12\Big)E_0-\frac1dm_0^2\phi_0^2-\Big(\frac3d-\frac32+1\Big)\frac1{3!}g_0\phi_0^3+\Big(\frac1d+\frac12\Big)j_0\phi_0\Big\}\\
&=\frac1{4(d-1)}\tmn\phi_0^2-\frac{d-2}{4(d-1)}\smn\phi_0^2-\frac1d\phi_0\tmn\phi_0\\
&\quad-\gmn\Big\{\Big(\frac1d-\frac12\Big)E_0-\frac1dm_0^2\phi_0^2+\frac{d/2-3}d\frac1{3!}g_0\phi_0^3+\Big(\frac1d+\frac12\Big)j_0\phi_0\Big\}.
\end{align*}

Now substitute for \(\phi_0^2, \phi_0\tmn\phi_0, \frac16g_0\phi_0^3\)
and \(j_0\phi_0\) from
\eqref{phi2vectoreq},\eqref{tensorrenorm},\eqref{3-matrix-eq}, and
\eqref{j_0-j}, i.e., repeating them here for convenience, 
\begin{align*}
\phi_0^2&=z_2[\phi^2]+\frac{4z_2m^2X}g\mu^{d/2-3}\phi+\frac{2C}{m_0^2}\partial^2\phi;\\
\phi_0\tmn\phi_0&=[\phi\tmn\phi]+\frac1{d-1}\Big\{\frac d4(z_2-1)-(d/2-3)a_7 - \xi_7\Big\}\tmn[\phi^2]\\
&\quad+ \frac1{d-1}\Big\{2C+dz_2m^2\mu^{d/2-3}-(d/2-3)a_4-\xi_4\Big\} \tmn\phi\\
&\quad+\frac1{d-1}\Big\{\frac{dC}{2m_0^2}-(d/2-3)a_6-\xi_6\Big\}\tmn\partial^2\phi;\\
\frac16g_0\phi_0^3&=\Big(1+\frac{\beta/g}{d/2-3}\Big)\frac16 g\mu^{3-d/2}[\phi^3]+\frac\delta{d/2-3}\frac12m[\phi^2]-\frac\gamma{d/2-3}[E]\\
&\quad-\Big(\frac\eta{d/2-3}+\frac{m^4X}{jg}\Big)j\mu^{d/2-3}\phi+a_4\partial^2\phi+a_6\partial^4\phi+a_7\partial^2[\phi^2];\\
j_0\phi_0&=(j+m^4X/g)\mu^{d/2-3}\phi.
\end{align*}

Thus,
\begin{align*}
T^C_{\mu\nu}&=\frac1{4(d-1)}\tmn\Big\{z_2[\phi^2]+\frac{4z_2m^2X}g\mu^{d/2-3}\phi+\frac{2C}{m_0^2}\partial^2\phi\Big\}+\frac{d-2}{4(d-1)}\smn\phi_0^2-\frac1d[\phi\tmn\phi]\\
&\quad-\frac1{d(d-1)}\bigg\{\Big\{\frac d4(z_2-1)-(d/2-3)a_7 - \xi_7\Big\}\tmn[\phi^2]\\
&\quad+ \Big\{2C+\frac{dz_2m^2X\mu^{d/2-3}}g-(d/2-3)a_4-\xi_4\Big\} \tmn\phi+\Big\{\frac{dC}{2m_0^2}-(d/2-3)a_6-\xi_6\Big\}\tmn\partial^2\phi\bigg\}\\
&\quad-\gmn\Big\{\Big(\frac{1-\gamma}d-\frac12\Big)[E]+\frac{\delta/2-1}dm^2[\phi^2]+\Big(\frac{1-\eta}d+\frac12\Big)j\mu^{d/2-3}\phi\\
&\quad+(d/2-3+\beta/g)\frac1{6d} g\mu^{3-d/2}[\phi^3] +\frac1d\{(d/2-3)a_4-2C\}\partial^2\phi+\frac{(d/2-3)}d\{a_6\partial^4\phi+a_7\partial^2[\phi^2]\}\Big\}.
\end{align*}
Again \(\because\)
\(-\gmn\partial^2=-\frac1{d-1}\tmn+\frac d{d-1}\smn\), (applying on $\partial^2\phi$),
\begin{align}
T^C_{\mu\nu}&=\frac{d/4+\xi_7}{d(d-1)}\tmn[\phi^2]+\frac{\xi_4}{d(d-1)}\tmn\phi+\frac{\xi_6}{d(d-1)}\tmn\partial^2\phi-\frac1d[\phi\tmn\phi]-\gmn\Big\{\Big(\frac{1-\gamma}d-\frac12\Big)[E]\\
&\quad+\frac{\delta/2-1}dm^2[\phi^2]+\Big(\frac{1-\eta}d+\frac12\Big)j\mu^{d/2-3}\phi+(d/2-3+\beta/g)\frac1{6d} g\mu^{3-d/2}[\phi^3]\Big\}\\
&\quad+\frac{d-2}{4(d-1)}\smn\phi_0^2+\frac{d/2-3}{d-1}a_7\smn[\phi^2]+\frac1{d-1}\{(d/2-3)a_4-2C\}\smn\phi+\frac{(d/2-3)}{d-1}a_6\smn\partial^2\phi,
\label{newEMTreno}
\end{align}
in which it can be seen that all the divergences are contained only in the terms in \eqref{newEMTreno}, in operators involving \(\smn\), and the other terms are all manifestly finite. Thus \(T_{\mu\nu}\) could be made finite.


Let us now define a new, improved energy momentum tensor that is finite:
\begin{equation}
\boxed{
\begin{aligned}
T_{\mu\nu}&\equiv T^C_{\mu\nu}-\frac1{d-1}(\partial_\mu\partial_\nu-\gmn\partial^2)\Bigg[\Big\{\frac{d-2}4+\frac{(d/2-3)a_7}{z_2}\Big\}\phi_0^2\\
&\quad+\frac1{\sqrt{z_1}}\Big\{(d/2-3)\Big(a_4-\frac4gm^2Xa_7\mu^{d/2-3}\Big)-2C\Big\}\phi_0+\frac1{\sqrt{z_1}}(d/2-3)\Big\{a_6-\frac{2C}{m^2}a_7\Big\}\partial^2\phi_0\Bigg],
\end{aligned}}
\label{finemt}
\end{equation}
or the manifestly finite expression,
\begin{align}
T_{\mu\nu}&=\frac{d/4+\xi_7}{d(d-1)}\tmn[\phi^2]+\frac{\xi_4}{d(d-1)}\tmn\phi+\frac{\xi_6}{d(d-1)}\tmn\partial^2\phi-\frac1d[\phi\tmn\phi]-\frac1d\gmn\Big\{\Big(1-\gamma-\frac d2\Big)[E]\\
&\quad+(\delta/2-1)m^2[\phi^2]+\Big(1-\eta+\frac d2\Big)j\mu^{d/2-3}\phi+(d/2-3+\beta/g)\frac16 g\mu^{3-d/2}[\phi^3]\Big\}\nonumber.
\end{align}

The divergence condition for \([\phi\tmn\phi]\)
\eqref{tensordivergence} gives
\begin{equation}
\partial^\nu T_{\mu\nu}=[E_\mu].
\end{equation}
The trace is obtained directly from the finite expression above,
\begin{align}
T_{\mu}^\mu=&-\Big(1-\gamma-\frac d2\Big)[E]+(\delta/2-1)m^2[\phi^2]+\Big(1-\eta+\frac d2\Big)j\mu^{d/2-3}\phi\nonumber\\
&+(d/2-3+\beta/g)\frac16 g\mu^{3-d/2}[\phi^3].
\end{align}
Thus the energy-momentum tensor is traceless if $m^2$, $j$ are set to zero and $g(d/2-3)+\beta(g)=0$, and the equations of motion are satisfied.

\subsection{Is \(T_{\mu\nu}\) RG invariant? }
The terms we have added to improve are not bare quantities. They involve terms that are dependent on renormalised parameters. From \eqref{newEMTreno},
\begin{align*}
T_{\mu\nu}&\equiv T^C_{\mu\nu}-\frac1{d-1}(\partial_\mu\partial_\nu-\gmn\partial^2)\bigg[\frac{d-2}{4}\phi_0^2+(d/2-3)a_7[\phi^2] +\{(d/2-3)a_4-2C\}\phi\\
&\quad+(d/2-3)a_6\partial^2\phi\bigg].
\end{align*}
Therefore,
\begin{align*}
\flow{} T_{\mu\nu}&=-\frac{d/2-3}{d-1}\Big(\flow {a_7}+a_7\flow{}\Big)\smn[\phi^2]-\frac1{d-1}\Big((d/2-3)\flow {a_4}-2\flow C\\
&\quad-(a_4(d/2-3)-2C)\gamma\Big)\smn\phi -\frac{d/2-3}{d-1}\Big(\flow{ a_6}-a_6\gamma\Big)s_{\mu\nu}\partial ^2\phi]          
\end{align*}
The required expressions for the RG derivatives are given by
\(\ref{a4flow},\ref{a6flow},\ref{a7flow}\), and \(\ref{Cflow}\), along
with
\begin{equation}
	\flow{[\phi^2]}=-\delta[\phi^2]-\frac{4j\mud}{m^2}(\gamma-\eta)\phi-2\zeta\mud\partial^2\phi,
	\label{{phi2flow}}
\end{equation}
This can be derived by applying the RG derivative to \eqref{phi2vectoreq}, and using the definition of the $\Gamma$ matrix \eqref{Gamma}.

Thus,
\begin{equation}
\begin{aligned}
\flow {T_{\mu\nu}}&=-\frac{1}{d-1}(d/2-3+\beta/g)\zeta_7\smn[\phi^2]\\
&\quad-\frac1{d-1}\Big((d/2-3+\beta/g)\zeta_4 +(\delta-2)\zeta\Big)m^2\mud\smn\phi\\
&\quad-\frac1{d-1}(d/2-3+\beta/g)\zeta_6\mud\smn\partial^2\phi.
\label{Tflow}
\end{aligned}
\end{equation}
Similar to equation (4.6) of \cite{brown_dimensional_1980}, this is completely transverse, finite, and it
is zero when $(d/2-3+\beta)$ and $m^2$ are zero at $d=6$, consistent with the requirements for a fixed point.

So the finite EM tensor we have constructed is not RG invariant. But it could be made RG invariant by adding a finite piece whose RG derivative cancels the finite pieces in \eqref{Tflow}\cite{collins_renormalization_1976}.

We begin by defining a new EM tensor by adding finite transverse pieces to $\emt$.
\begin{equation}
	\nemt\equiv\emt+b_4m^2\mud\smn\phi+b_6\mud\smn\psq\phi+b_7\smn[\phi^2],
	\label{invemt}
\end{equation}
where $b_4$, $b_6$, $b_7$ are finite functions of $g$, (they are dimensionless), such that 
\begin{equation}
\flow{\nemt}=O(d-6).
\label{nemtcondition}
\end{equation}
Note that we are requiring it to be $O(d-6)$ and not $0$.
From \eqref{{phi2flow}} and
\begin{equation}
\flow{\phi}=-\gamma\phi,
\end{equation}
we have,
\begin{align}
\flow{\nemt}
&=\flow{\emt} +\bigg(\flow{b_4} +(d/2-3+\delta)b_4\bigg)m^2\mud\smn\phi\nonumber\\
&\quad +\bigg(\flow{b_6} +(d/2-3)b_6\bigg)\mud\smn\psq\phi +\flow{b_7}\smn[\phi^2]\nonumber\\
&\quad -\gamma b_4m^2\mud\smn\phi -\gamma b_6\mud\smn\psq\phi \nonumber\\
&\quad -b_7\{\delta\smn[\phi^2] -4m^2\mud\bar{\eta}\smn\phi +2\zeta\mud\smn\psq\phi\}, 
\end{align}
where we have substituted $-m^4\bar{\eta}/j $ for $\gamma-\eta$ as in \eqref{bar_eta}.

After substituting for $\flow{\emt}$ from \eqref{Tflow}, we look at the coefficients of each of the linearly independent operators.

The coefficient of $\smn[\phi^2]$ is
\begin{equation}
	\flow{b_7} -\delta b_7 -\frac1{d-1}(d/2-3+\beta/g)\zeta_7. 
\end{equation}
We can replace the $\flow{}$ with $(d/2-3+\beta/g)g\gder{}$ because $b_4$ is a function of $g$ alone.

If we demanded that this coefficient is 0, along with the coefficients for the other two operators, $\flow{\nemt}$ would be zero, but then solving for $b_7$ wouldn't give a finite result. To avoid this, we drop the terms proportional to $(d-6)$  and find a $b_7$ that satisfies, and this is why we are able to obtain \eqref{nemtcondition} only to $O(d-6)$.
\begin{equation}
	-\frac1{d-1}\frac{\beta}g \zeta_7+\beta\gder{b_7}-\delta b_7=0.
	\label{b7key}
\end{equation}
This gives
\begin{equation}
	b_7(g) =\frac1{d-1}e^{\int_{g^*}^{g}\frac{\delta}{\beta}\dd g'}\int_{g^*}^{g}\frac{\dd g'}{g'}\zeta_7(g')e^{-\int_{g^*}^{g'}\frac{\delta}{\beta}\dd g''},
\end{equation}
which is finite. Here we have set the free integration limit to fixed point value $g^*$ to ensure the $\nemt$ is the same as $\emt$ at the fixed point, where it is traceless. 

The coefficient of $m^2\mud\smn\phi$ is
\begin{equation}
	\flow{b_4} +(d/2-3+\delta-\gamma)b_4 +4\bar{\eta} b_7 -\frac{1}{d-1}\{(d/2-3+\beta/g)\zeta_4 +(\delta-2)\zeta\}. 	
\end{equation}
$b_4$ then has to satisfy
\begin{equation}
	\beta\gder{b_4} +(\delta-\gamma)b_4 +4\bar{\eta} b_7 -\frac{1}{d-1}\{\beta\zeta_4/g +(\delta-2)\zeta\}=0,	
	\label{b4key}
\end{equation}
giving
\begin{equation}
	b_4(g) =\frac1{d-1}e^{-\int_{g^*}^{g}\frac{\delta-\gamma}{\beta}\dd g'} \int_{g^*}^{g}\{\beta\zeta_4/g' +(\delta-2)\zeta -4(d-1)\bar{\eta}b_7\} e^{\int_{g^*}^{g'}\frac{\delta-\gamma}{\beta}\dd g''} \dd g'. 
\end{equation}

Finally, the coefficient of $\mud\smn\psq\phi$ is
\begin{equation}
	\flow{b_6} +(d/2-3-\gamma) b_6 +2b_7\zeta-\frac1{d-1}(d/2-3+\beta/g)\zeta_6.
\end{equation}
$b_6$ will have to satisfy
\begin{equation}
\beta\gder{b_6}-\gamma b_6 +2b_7\zeta-\frac1{d-1}\beta\zeta_6/g=0,
\label{b6key}
\end{equation}
giving
\begin{equation}
	b_6=\frac1{d-1}e^{\int_{g^*}^{g}\frac{\gamma}{\beta}\dd g'}\int_{g^*}^{g}\dd g'(\beta\zeta_6/g'-2(d-1)\zeta b_7) e^{-\int_{g^*}^{g'}\frac{\gamma}{\beta}\dd g''}.
\end{equation}

The RG derivative of $\nemt$ is then
\begin{align}
	\flow{\nemt}= &-\frac{d/2-3}{d-1}\{\zeta_7\smn[\phi^2] +m^2\zeta_4\mud\smn\phi +\zeta_6\mud\smn\psq\phi\} \nonumber\\
	&\quad +(d/2-3)m^2\mud(b_4+\gder{b_4})\smn\phi +(d/2-3)\mud(b_6+\gder{b_6})\smn\psq\phi\nonumber\\
	&\quad +(d/2-3)\gder{b_7}\smn[\phi^2], 
\end{align}
or from \eqref{b7key}, \eqref{b4key}, \eqref{b6key},
\begin{align}
	\flow{\nemt} = &(d/2-3)\frac{gm^2\mud}{\beta}\bigg\{(\beta/g+\gamma-\delta)b_4 -4\bar{\eta}b_7 +\frac{1}{d-1}\zeta
	(\delta-2) \bigg\}\smn\phi\nonumber\\
	&+(d/2-3)\frac{g\mud}{\beta}\{(\beta/g+\gamma) -2\zeta b_7 \} \smn\psq\phi\nonumber\\
	&+(d/2-3) \frac{g\delta b_7}{\beta}\smn[\phi^2].
  \label{nemtflow}
\end{align}
This clearly vanishes at $d=6$.

We can write down the expression for $\nemt$ from \eqref{finemt} and \eqref{invemt}, also substituting for $[\phi^2]$ with bare quantities from \eqref{phi2vectoreq},
\begin{align}
	\nemt= & T^C_{\mu\nu}-\frac{d-2}{4(d-1)}\smn\phi_0^2 +\frac1{z_2}\Big(b_7-\frac{d/2-3}{d-1}a_7\Big)\smn\phi_0^2\nonumber\\
	&+\frac1{\sqrt{z_1}}\Big\{\Big(m^2\mud b_4-\frac{d/2-3}{d-1}a_4+\frac{2C}{d-1}\Big)-\frac4gm^2X\mud\Big(b_7-\frac{d/2-3}{d-1}a_7\Big)\Big\}\smn\phi_0\nonumber\\
	&+\frac1{\sqrt{z_1}}\Big\{b_6\mud -\frac{d/2-3}{d-1}a_6-\frac{2C}{m^2}\Big(b_7-\frac{d/2-3}{d-1}a_7\Big)\Big\}\smn\psq\phi_0.
  \label{nemt}
\end{align}

\section{Concluding Remarks}
We have used dimensional regularisation and the action principle to derive the operator mixing in $\phi^3$ theory in six dimensions. The matrix has a similar structure to the one obtained for $\phi^4$ theory in 4 dimensions\cite{brown_dimensional_1980}, but it had been made more complicated by the fact that there are three operators with total derivatives that $\phi^3$ can mix with, and one operator with total derivative that $\phi^2$ can mix with. The four new anomalous dimensions that are obtained are associated with composite operators, and are calculated by renormalising correlation functions with composite operator insertions in appendix \ref{oneloopreno}. They all turn out to be zero at one loop order.

The tensor operator renormalisation similarly is more complicated the $\phi^4$ case, but essentially the same procedure. The EM tensor renormalisation requires subtracting transverse terms involving three operators. Thus there must be three non-minimal couplings in the action for the theory when weak gravitation is taken into account.

The divergence of the EM tensor is given by the equation of motion operator $[E_\mu]$ as mentioned in the introduction. The trace of the minimally subtracted $\emt$ has anomalous terms and conformal invariance is thus absent. The trace vanishes at fixed point onshell. 

We check for RG invariance and find that the RG derivative of $\emt$ is proportional to the $\beta$ function and the new anomalous dimensions corresponding to operator mixing, but there is also one additional term proportional to $(\delta-2)$ and the mass parameter. Yet this vanishes at the fixed point, as it should.

The RG invariant $\nemt$ also obeys the same properties, viz., its divergence is zero, trace vanishes at fixed point, and the RG derivative is zero at fixed point. But, as can be seen from \eqref{nemtflow}, $\nemt$ is RG coinvariant only at $d=6$. Further, the RG derivative of $\nemt$ diverges if $\beta/g=0$, but this is only expected to happen for either $g=0$ (the free theory where these are not required anyway), or for some $g\sim O(1)$ (in which case the perturbation theory is not valid).



In dimensional regularization the renormalization procedure involves an expansion in powers $\frac{1}{\epsilon}$\cite{t_hooft_dimensional_1973}. It has been suggested there that it may be possible to resum this series and obtain non-perturbative results. In \cite{collins_renormalization_1976} this has been attempted for the energy momentum tensor improvement terms  for the $\phi^4$ theory in $D=4$. It would be interesting to explore this idea further and also apply it to the theory studied in this paper.

It would also be interesting to construct the fixed point actions and the energy momentum tensor for theories in $D=6\pm \epsilon$ using the exact Renormalization Group equations as has been done recently for the Wilson-Fisher fixed point theory in $D=4-\epsilon$\cite{Dutta:2020vqo}.

\appendix
\section{One-loop Counterterms for the Lagrangian and RG functions}
\label{appendix:oneloop}
We will evaluate the one-loop diagrams that contribute to the counterterms in the Lagrangian.
We will need this expansion in someplaces below.
\begin{equation}
\lim\limits_{\epsilon\to0}\Gamma(-n+\epsilon)=\frac{(-1)^n}{n!}\big[\frac1\epsilon+\psi_1(n+1)+O(\epsilon)\big]\text{, for }n\geq0.
\label{Gamma_Poles}
\end{equation}

\subsection{Mass and wave-function renormalisation}
\label{masswaverenorm}
The diagram that contributes is 
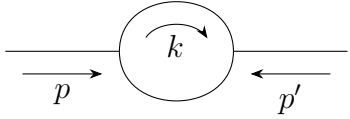
\begin{figure}[ht]
	\centering
	\begin{tikzpicture}
	\begin{feynman}
	\vertex (i);
	\vertex[right= of i] (a);
	\vertex[right= of a] (b);
	\vertex[right= of b] (f);
	\diagram* {
		(i) -- [momentum'=$p$] (a)
		-- [half left, momentum'=$k$] (b)
		-- [half left] (a),
		(f) -- [momentum=$p^\prime$] (b)
	};
	\end{feynman}
	\end{tikzpicture}
	\caption{one loop self-energy diagram}
\end{figure}
\begin{align*}
\Sigma(p,p^\prime)/i&\equiv\Sigma(p)\delta(p+p^\prime)/i=\Pi_2(p)\\
&=(-ig)^2\frac{\mu^{6-d}}{(2\pi)^{2d}}\times\frac{3\times3\times2}{6\times6}\int\limits_{k_1}\int\limits_{k_2}\mathrm{d}^dk_1\ \mathrm{d}^dk_2\ \Delta(k_1)\Delta(k_2)\ \delta(p+k_1+k_2)\delta(p^\prime+k_1+k_2)\\
&=\frac12(-ig)^2\frac{\mu^{6-d}}{(2\pi)^d}\int\limits_k\mathrm{d}^dk\ \Delta(k)\Delta(p-k)\delta(p+p^\prime)
\end{align*}
\begin{align*}
\Sigma(p)/i&=A\int\limits_k\mathrm{d}^dk\ \frac{i}{k^2-m^2+i\epsilon}\frac{i}{(k-p)^2-m^2+i\epsilon}\\
&=A\int_0^\infty\mathrm{d}a\int_0^\infty\mathrm{d}b\int\limits_k\mathrm{d}^dk\ \exp i\{a(k^2-m^2+i\epsilon) + b((k-p)^2-m^2+i\epsilon)\}\\
&=A\int\ \exp i\{(a+b)k^2-(a+b)m^2 + bp^2-2bk.p+i(a+b)\epsilon)\}. \\
\end{align*}
The second equation is after using \emph{Schwinger's Trick}\cite{wilson_quantum_1973}.
\begin{align*}
\text{Shift }k_\mu\rightarrow k_\mu-\frac{bp_\mu}{a+b};\ \text{then }z\equiv a+b;\ x\equiv \frac b{a+b};\ \mathrm{d}a\mathrm{d}b=z\mathrm{d}z\mathrm{d}x,\\
\Sigma/i=A\int\mathrm{d}^dk\int_0^\infty\mathrm{d}z\int_0^1\mathrm{d}x\ z\exp iz\{k^2-(m^2-p^2x(1-x))+i\epsilon\},\\
\int \mathrm{d}^dk\ e^{izk^2}=z^{-\frac d2}\int \mathrm{d}^dk\ e^{ik^2}=z^{-\frac d2}\pi^{\frac d2}i^{1-\frac d2},
\end{align*}
\begin{align*}
\Sigma/i&=A\times\pi^{\frac d2}i^{1-\frac d2}\int_0^1\mathrm{d}x\int_0^\infty\mathrm{d}z\ z^{1-\frac d2}\exp (-iz\Delta)\\
&=A\times\pi^{\frac d2}i^{1-\frac d2}\int_0^1\mathrm{d}x\ \Gamma(2-d/2)(i\Delta)^{\frac d2-2}\\
&=\frac12(-ig)^2\frac{\mu^{6-d}}{(2\pi)^d}\times \pi^{\frac d2}\times(-i)\times\Gamma(2-d/2)\int_0^1\mathrm{d}x\ [m^2-p^2x(1-x)]^{\frac d2-2}\\
&=i\frac{g^2}{2(4\pi)^3}\bigg(-\frac{1}{3-d/2}+finite\bigg)\int_0^1\mathrm{d}x\ (m^2-p^2x(1-x))\bigg[\frac{m^2-p^2x(1-x)}{4\pi\mu^2}\bigg]^{\frac d2-3}\\
&=-i\frac{g^2}{2(4\pi)^3(3-d/2)}\int_0^1\mathrm{d}x\ (m^2-p^2x(1-x))+finite\\
&=-i\frac{g^2}{(4\pi)^3(6-d)}(m^2-p^2/6)+finite;\\
\Sigma&=\frac{g^2}{(4\pi)^3(6-d)}(m^2-p^2/6).
\end{align*}
To subtract this, we need the following counter-terms for the parameters.
\begin{equation}
\boxed{
	\begin{aligned}
	\therefore \delta m^2&=\frac{g^2m^2}{64\pi^3(d-6)}\\
	\delta z_1&=\frac{g^2}{6\times64\pi^3(d-6)}
	\end{aligned}
}
\label{massCT}
\end{equation}
Refer the counter-term Lagrangian in \eqref{CTLag}.

\subsection{Vertex renormalisation}
\label{vertexrenorm}
This is for the vertex diagram at one loop.
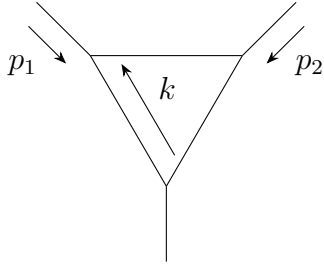
\begin{figure}[ht]
	\centering
	\begin{tikzpicture}
	\begin{feynman}
	\vertex (i) at (0,0);
	\vertex (a) at (0.7071,-0.7071);
	\vertex (c) at (1.7071,-2.4392);
	\vertex (b) at (2.7071,-0.7071);
	\vertex (j) at (1.7071,-3.4392);
	\vertex (k) at (3.414,0);
	\diagram* {
		(i) -- [momentum'=$p_1$] (a)
		-- (b)
		-- (c)
		-- [momentum'=$k$](a),
		(j) -- (c),
		(k) -- [momentum=$p_2$] (b)
	};
	\end{feynman}
	\end{tikzpicture}
	\caption{one loop vertex diagram}
\end{figure}
\begin{align*}
\Gamma(p_1,p_2)&=\frac{ig^3(\mu^{9-3d/2})}{(2\pi)^d}\int\mathrm{d}^dk\ \frac{i}{k^2-m^2+i\epsilon}\frac{i}{(k+p_1)^2-m^2+i\epsilon}\frac{i}{(k-p_2)^2-m^2+i\epsilon}\\
&=A(-i)^3\int\mathrm{d}^dk\int_0^\infty\mathrm{d}a\int_0^\infty\mathrm{d}b\int_0^\infty\mathrm{d}c \ \exp i\{a(k^2-m^2+i\epsilon) + b((k+p_1)^2-m^2+i\epsilon)\\
&\quad+ c((k-p_2)^2-m^2+i\epsilon)\} \\
&=(\dots)\int\ \exp i\{(a+b+c)k^2-(a+b+c)m^2 + bp_1^2+cp_2^2+2bk.p_1-2ck.p_2+i(a+b+c)\epsilon)\} \\
&=(\dots)\int\ \exp i\{(a+b+c)k^2-(a+b+c)m^2 + bp_1^2+cp_2^2+\frac{(bp_1-cp_2)^2}{a+b+c}+i(a+b+c)\epsilon)\} \\
\end{align*}
\begin{gather*}
r=a+b+c;\ x=\frac br;\ y=\frac cr;\ z=\frac ar;\\
\mathrm{d}a\mathrm{d}b\mathrm{d}c=r^2\mathrm{d}r\mathrm{d}x\mathrm{d}y\mathrm{d}z\ \delta(x+y+z-1),\nonumber\\
\Gamma(p_1,p_2)=A(i)^3\int\mathrm{d}^dk\int_0^\infty\mathrm{d}r\int_0^1\mathrm{d}x\int_0^{1-x}\mathrm{d}y\\
r^2\exp i\{rk^2-rm^2+rxp_1^2+ryp_2^2-r(xp_1-yp_2)^2+ir\epsilon\},\\
\text{Scale }k\rightarrow k\sqrt{r},\\
\int \mathrm{d}^dk\ e^{irk^2}=r^{-\frac d2}\int \mathrm{d}^dk\ e^{ik^2}=r^{-\frac d2}\pi^{\frac d2}(-i)^\frac12i^\frac{d-1}{2},\\
\Gamma(p_1,p_2)=A(-i)^{4-\frac d2}\int_0^1\mathrm{d}x\int_0^{1-x}\mathrm{d}y\int_0^\infty\mathrm{d}r\ r^{2-\frac d2}\exp(-ir\Delta)\times\pi^{\frac d2},\\
\text{where }\Delta=m^2-xp_1^2-yp_2^2+(xp_1-yp_2)^2;\\
\text{Scale } r\rightarrow ir\Delta;\ \int_0^{i\infty}\mathrm{d}r\ r^z e^{-r}=\Gamma(z+1),
\end{gather*}
\begin{align*}
\Gamma(p_1,p_2)&=A(-i)^{4-\frac d2}\times\pi^{\frac d2}\times\Gamma(3-d/2)\int_0^1\mathrm{d}x\int_0^{1-x}\mathrm{d}y\ \frac{1}{i\Delta}\times (i\Delta)^{\frac d2-2}\\
&=\frac{g^3(\mu^{9-3d/2})}{(2\pi)^d}\times(-i)^{7-d}\times\pi^{\frac d2}\times\Gamma(3-d/2)\int_0^1\mathrm{d}x\int_0^{1-x}\mathrm{d}y\ \Delta^{\frac d2-3}\\
&=\frac{-ig^3(\mu^{3-d/2})}{(4\pi)^3}\times\Gamma(3-d/2)\int_0^1\mathrm{d}x\int_0^{1-x}\mathrm{d}y\ \bigg[\frac{m^2-xp_1^2-yp_2^2+(xp_1-yp_2)^2}{4\pi\mu^2}\bigg]^{\frac d2-3}\\
&=\frac{-ig^3(\mu^{3-d/2})}{(4\pi)^3}\times\bigg[\frac{1}{3-d/2}+const.\bigg]\times\bigg[\frac 12+\Big(\frac d2-3\Big)(\dots)+\ldots\bigg]\\
&=\frac{-ig^3(\mu^{3-d/2})}{(4\pi)^3(6-d)}+finite.
\end{align*}
The required counter-term to cancel this is
\begin{equation}
\boxed{\therefore \delta g=\frac{g^3}{(4\pi)^3(d-6)}}+O(g^5)
\label{vertexcounterterm}
\end{equation}

\subsection{Source term renormalisation}

To find the counterterm for the linear coupling, we compute the tadpole diagram.
\begin{figure}[ht]
	\centering
	\begin{tikzpicture}
		\begin{feynman}
		\vertex (i);
		\vertex[above = 1cm of i] (a);
		\vertex[above = of a] (b);
		\diagram*{
		(i) -- [momentum=$p$](a)
		-- [half left, momentum'=$k$](b)
		-- [half left](a) 
		};
		\end{feynman}
	\end{tikzpicture}
	\caption{Tadpole diagram}
\end{figure}
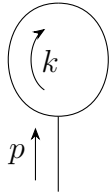

	\(j\int\mathrm{d}^dx\ \phi(x)=j\int\mathrm{d}^dx\int\mathrm{d}^dp\ \tilde{\phi}(p)e^{ip.x}=j\int\mathrm{d}^dp\ \tilde{\phi}(p)\delta(p)\).
	This is the term in momentum description.
\begin{align*}
I&=-\frac{3\times ig\mu^{3-d/2}}{3!(2\pi)^d}\int\mathrm{d}^dk\ \frac{i}{k^2-m^2+i\epsilon}\\
&=-\frac{ig\mu^{3-d/2}}{2(2\pi)^d}\int\mathrm{d}^dk\int_0^\infty\mathrm{d}a\ \exp i\{a(k^2-m^2+i\epsilon)\}\\
&=(\dots)\int_0^\infty\mathrm{d}a\ a^{-d/2}\int\mathrm{d}^dk\ \exp i\{(k^2-am^2+i\epsilon)\}\\
&=(\dots)\int_0^\infty\mathrm{d}a\ a^{-d/2}e^{-iam^2}\times\pi^{d/2}\times i^{d/2-1},
\end{align*}
\begin{equation}
a\rightarrow iam^2;\ \int_0^{i\infty}\mathrm{d}a\ a^ze^{-a}=\Gamma(z+1),\nonumber
\end{equation}
\begin{align*}
I&=-\frac{ig\mu^{3-d/2}}{2(2\pi)^d}\times\pi^{d/2}\times(im^2)^{d/2-1}\times\Gamma(1-d/2)\\
&=-\frac{i^{d-1}g\mu^{3-d/2}(m^2)^{d/2-1}}{2(4\pi)^{d/2}}\Gamma(1-d/2)\\
&=-\frac{i^5gm^4\mu^{d/2-3}}{2(4\pi)^3}\times\frac12\bigg[\frac1{3-d/2}+finite\bigg]\times\bigg[1+(d/2-3)\log\bigg(\frac{m^2}{\mu^2}\bigg)+\dots\bigg]\\
&=-\frac{igm^4\mu^{d/2-3}}{2(4\pi)^3(6-d)}.
\end{align*}
\begin{equation}
\boxed{\therefore \delta j=-\frac{gm^4}{2(4\pi)^3(d-6)}}+O(g^3).
\label{deltaj}
\end{equation}
It does not depend on \(j\).

\subsection{RG functions}

\subsubsection{\(\beta\)-function}
\begin{align}
g_0\phi_0^3&=(g+\delta g)\phi^3\mu^{3-d/2}\nonumber\\
g_0(1+\delta z_1)^{3/2}&=(g+\delta g)\mu^{3-d/2}\nonumber\\
g_0\bigg(1+\frac{g^2}{6\times64\pi^3(d-6)}\bigg)^{3/2}&=g\bigg(1+\frac{g^2}{64\pi^3(d-6)}\bigg)\mu^{3-d/2}\nonumber\\
g_0&=g\bigg(1+\frac{g^2}{64\pi^3(d-6)}\bigg)\bigg(1+\frac{g^2}{6\times64\pi^3(d-6)}\bigg)^{-3/2}\mu^{3-d/2}\nonumber\\
&\cong g\bigg(1+\bigg(1-\frac32\times\frac16\bigg)\frac{g^2}{64\pi^3(d-6)}\bigg)\mu^{3-d/2}\nonumber\\
g_0&\cong g\bigg(1+\frac34\frac{g^2}{64\pi^3(d-6)}\bigg)\mu^{3-d/2}+O(g^5).\\
\end{align}
\begin{gather*}
\mu\frac{\partial g_0}{\partial \mu}=0\\
\implies0=\mu\frac{\partial g}{\partial \mu}\bigg(1+\frac34\frac{3g^2}{64\pi^3(d-6)}\bigg)\mu^{3-d/2}+g\bigg(1+\frac34\frac{g^2}{64\pi^3(d-6)}\bigg)(3-d/2)\mu^{3-d/2}\\
\mu\frac{\partial g}{\partial \mu}=-g(3-d/2)\bigg(1+\frac34\frac{g^2}{64\pi^3(d-6)}\bigg)\bigg(1+\frac34\frac{3g^2}{64\pi^3(d-6)}\bigg)^{-1}\\
\mu\frac{\partial g}{\partial \mu}=g(d/2-3)-\frac{3g^3}{256\pi^3}+O(g^5).
\end{gather*}
\begin{equation}
\boxed{\therefore \beta(g)\equiv\mu\frac{\partial g}{\partial \mu}-g(d/2-3)=-\frac{3g^3}{256\pi^3}}+O(g^5).
\end{equation}

\subsubsection{Anomalous dimension}
\begin{align}
\gamma&\equiv\frac12\mu\frac{\mathrm d}{\mathrm d\mu}\log z_1\nonumber\\
&=\frac12\mu\frac{\partial g}{\partial \mu}\frac{\partial}{\partial g}\log z_1\nonumber\\
&=\frac12\bigg(g(d/2-3)-\frac{3g^3}{256\pi^3}\bigg)\times\frac{\partial}{\partial g}\bigg(\frac{g^2}{6\times64\pi^3(d-6)}+higher\ poles\bigg)\nonumber\\
&=\frac12g\times(d/2-3)\times\frac{2g}{6\times64\pi^3(d-6)}+\dots\nonumber
\end{align}
\begin{equation}
\boxed{
\gamma(g)=\frac{g^2}{12(4\pi)^3}}+O(g^3).
\end{equation}
\subsubsection{Mass anomalous dimension}
\begin{align*}
m_0^2\phi_0^2&=m^2\phi^2+\delta m^2\phi^2\\
&=(m^2+\delta m^2)\frac{\phi_0^2}{z_1}\\
\implies m_0^2&=\bigg(1+\frac{g^2}{64\pi^3(d-6)}\bigg)\bigg(1+\frac{g^2}{6\times64\pi^3(d-6)}\bigg)^{-1}\times m^2\\
&=\bigg(1+\frac{5g^2}{6\times64\pi^3(d-6)}+O(g^4)\bigg)m^2.\\
\implies\mu\frac{\mathrm dm_0^2}{\mathrm d\mu}&=\mu\frac{\mathrm dm^2}{\mathrm d\mu}\bigg(1+\frac{5g^2}{384\pi^3(d-6)}\bigg)+m^2\times\big(\beta+g(d/2-3)\big)\bigg(\frac{5g}{192\pi^3(d-6)}+O(g^3)\bigg).
\end{align*}
\begin{equation}
\boxed{\therefore \delta(g)\equiv\mu\frac{\mathrm d}{\mathrm d\mu}\log m^2=-\frac{5g^2}{384\pi^3}}+O(g^4).
\end{equation}

\subsubsection{RG of \(j\)}
From the form of \(j_0\),
\begin{align*}
j_0(1+\delta z_1)^\frac12&=(j+\delta j)\mu^{d/2-3}\\
j_0&=(1+\delta z_1)^{-\frac12}(j+\delta j)\mu^{d/2-3}\\
\text{differentiating w.r.t. }\mu,\\
0&=-\frac12(1+\delta z_1)^{-\frac32}(j+\delta j)\mu^{d/2-3}\mu\frac{\mathrm d(\delta z_1)}{\mathrm d\mu}+(1+\delta z_1)^{-\frac12}\mu\frac{\mathrm d}{\mathrm d\mu}(j+\delta j)\mu^{d/2-3}\\
&\quad+(d/2-3)(1+\delta z_1)^{-\frac12}(j+\delta j)\mu^{d/2-3}\\
0&=-\frac12(1+\delta z_1)^{-1}(j+\delta j)\mu\frac{\mathrm d(\delta z_1)}{\mathrm d\mu}+\mu\frac{\mathrm d}{\mathrm d\mu}(j+\delta j)+(d/2-3)(j+\delta j)\\
\mu\frac{\mathrm dj}{\mathrm d\mu}&=-(d/2-3)(j+\delta j)-\mu\frac{\mathrm d(\delta j)}{\mathrm d\mu}+\frac12(1+\delta z_1)^{-1}(j+\delta j)\mu\frac{\mathrm d(\delta z_1)}{\mathrm d\mu}\\
\mu\frac{\mathrm dj}{\mathrm d\mu}&=-(d/2-3)(j+\delta j)-\frac{\delta j}{g}\mu\frac{\mathrm dg}{\mathrm d\mu}-2\delta j\delta(g)\\
&\quad +\frac12(1+O(g^2))^{-1}(j+\delta j)\frac{2g}{6\times64\pi^3(d-6)}\mu\frac{\mathrm dg}{\mathrm d\mu}\\
\mu\frac{\mathrm dj}{\mathrm d\mu}&=-(d/2-3)(j+\delta j)-\delta j\big((d/2-3)+O(g^2)\big)-\delta j\times O(g^2)\\
&\quad+\frac12(1+O(g^2))^{-1}(j+\delta j)\frac{2g}{6\times64\pi^3(d-6)}\big(g(d/2-3)+O(g^3)\big)\\
\mu\frac{\mathrm dj}{\mathrm d\mu}&=-(d/2-3)j+\frac{gm^4}{2\times 64\pi^3}+\frac{jg^2}{12\times64\pi^3}+O(g^3)\\
\end{align*}
Define a RG function \(\eta(g,m^4/j)\), where
\(\mu\frac{\mathrm dj}{\mathrm d\mu}\equiv j\eta-(d/2-3)j\).
\begin{equation}
\text{Thus, }\boxed{\eta(g,m^4/j)=\frac{g}{2(4\pi)^3}\bigg(\frac g6+\frac{m^4}{j}\bigg)}+O(g^3, g^2(m^4/j)^2).
\end{equation}

\section{One-loop Renormalization of Composite Operators}
\label{oneloopreno}
We verify the results of $\phi^2$ and $\phi^3$ renormalisation at least to leading order, by comparing it with the result from computing Feynman diagrams. 
The $\phi^2$ renormalisation has been done in \cite{collins_1984}To renormalise $\phi^2$ we look at the following correlation function.
\begin{equation}
\langle 0|T\phi(x)\phi(y)\phi^2(z)/2|0\rangle.
\end{equation}
To order $g^2$, the graphs with $\phi^2$ insertions, (insertion shown with a cross in the figure below), that need renormalisation are
\begin{figure*}[h]
	\centering
	\begin{subfigure}[b]{0.1\textwidth}
		\begin{tikzpicture}
		\begin{feynman}
		\vertex (i);
		\vertex[right= of i] (f);
		\diagram*{
			(i) -- [insertion=0.5] (f)
		};
		\end{feynman}
		\end{tikzpicture}
		
		\vspace*{0.75cm}
		\caption{}
		\label{phi2tree}
	\end{subfigure}
	\begin{subfigure}[b]{0.2\textwidth}
	\begin{tikzpicture}
	\begin{feynman}
	\vertex (i);
	\vertex[right= 1cm of i] (a);
	\vertex[right= of a] (b);
	\vertex[right= 1cm of b] (f);
	\diagram* {
		(i) -- (a)
		-- [half left, insertion=0.5] (b)
		-- [half left] (a),
		(f) -- (b)
	};
	\end{feynman}
	\end{tikzpicture}
	\caption{}
	\label{phi2selfenergy}
	\end{subfigure}
	\begin{subfigure}[b]{0.13\textwidth}
		\begin{tikzpicture}
		\begin{feynman}
		\vertex (i);
		\vertex[right= 1cm of i] (a);
		\vertex[right= 1cm of a] (f);
		\vertex[above= 0.5cm of a] (b);
		\vertex[above= of b] (c);
		\diagram*{
			(i) -- (f),
			(a) -- (b)
			-- [half left, insertion=0.99](c) 
			-- [half left](b) 
		};
		\end{feynman}
		\end{tikzpicture}
		\caption{}
		\label{phi2tadpole}
	\end{subfigure}

\caption{Renormalisation of $\phi^2$ to one-loop}
\end{figure*}
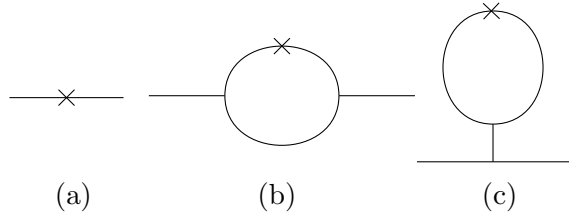

In momentum space, the lowest-order graph \ref{phi2tree} is
\begin{equation}
	G_a= \Delta(p_1)\Delta(p_2),
\end{equation}  
where \(\Delta(p)\) is the Feynman propagator,
\[\Delta(p)=\frac i{p^2-m^2+i\epsilon}.\]
For fig. \ref{phi2selfenergy},
\begin{align}
	G_b= &\Delta(p_1)\Delta(p_2)\times\bigg\{-\frac{g^2\mu^{6-d}}{(2\pi)^d}\int \mathrm d^dk\, \Delta(k)\Delta(k-p_1)\Delta(k+p_2)\bigg\}\nonumber\\
	=&\Delta(p_1)\Delta(p_2)\times\frac{g^2}{64\pi^3(6-d)}+finite\ldots
\end{align}
where we have written down the result from the integral computed in \ref{vertexrenorm}.

For fig. \ref{phi2tadpole},
\begin{align}
G_c=&ig\mu^{3-d/2}\Delta(p_1)\Delta(p_2)\Delta(p_1+p_2)\times \frac{ig\mu{3-d/2}}{2(2\pi)^d}\int\mathrm d^dk\, \Delta(k)\Delta(p_1+p_2+k)\nonumber\\
=&ig\mu^{3-d/2}\Delta(p_1)\Delta(p_2)\Delta(p_1+p_2)\times\bigg\{\frac{g\mu^{3-d/2}}{64\pi^3(6-d)}[m^2-\frac16(p_1+p_2)^2]\bigg\}+finite\ldots
\end{align}
This result is from the computation in \ref{masswaverenorm}.

To subtract these divergences, we look at the kind of vertices that can replace the loops in \ref{phi2selfenergy} and \ref{phi2tadpole}. These are $\phi^2$,and $\phi$ respectively.

Thus, the renormalised
\begin{equation}
\frac12[\phi^2]\equiv\bigg[1+\frac{g^2}{64\pi^3(d-6)}\bigg]\frac12\phi^2+\frac{g\mu^{d/2-3}}{64\pi^3(d-6)}(m^2+\frac16\partial^2)\phi+higher\ orders
\label{collinsResult}
\end{equation}  
We obtained the result to be compared in \eqref{phi2vectoreq}:
\begin{align}
\frac12m^2[\phi^2] &= \frac12m_0^2\phi_0^2 - Bj\mud\phi - C\partial^2\phi\nonumber\\
\frac12[\phi^2] &= \frac{m_0^2z_1}{m^2}\frac12\phi^2 -\frac{Bj\mud}{m^2}\phi - \frac{C}{m^2}\partial^2\phi \label{phi2renorm}
\end{align}
For the coefficients, recall \(m_0^2z_1 = m^2+\delta m^2\) and \(B=2\delta j/j\) \eqref{phi2vectoreqresult}. We get the
values of \(\delta m^2\) and \(\delta j\) from \eqref{massCT} and
\eqref{deltaj}.
\begin{align}
\frac12[\phi^2] &= \bigg(1+\frac{\delta m^2}{m^2}\bigg)\frac12\phi^2 -\frac{2\delta j}{m^2}\phi-\frac{C}{m^2}\partial^2\phi \nonumber\\
&= \bigg(1+\frac{g^2}{(4\pi)^3(d-6)}\bigg)\frac12 \phi^2 + \frac{gm^2}{(4\pi)^3(d-6)}\phi - \frac{C}{m^2}\partial^2\phi.
\end{align}
So coefficients of $\phi^2$ and \(\phi\) match.

We can also obtain the value for $C$ to one loop from this comparison.
\begin{equation}
	\frac12 \phi^2=\bigg(1-\frac{g^2}{64\pi^3(d-6)}\bigg)\frac12[\phi^2]-\frac{g\mud}{64\pi^3(d-6)}(m^2+\frac16\Box)\phi+higher orders.
\end{equation}
Comparing this with \eqref{phi2renorm}, we get
\begin{equation}
	C=-\pole[6\times]{m^2g\mud}+ O(g^2).
\end{equation}
\eqref{Cflow} then gives $\zeta$; it is just the finite part of $\flow{C}$.
\begin{equation}
	\flow{C}=-\frac{g\mud m^2}{12\times 64\pi^3}-\frac{g\mud m^2}{12\times 64\pi^3}+divergent+O(g^2).
\end{equation}
Thus,
\begin{equation}
	\zeta(g)=-\frac{g}{6\times 64\pi^3}+O(g^2).
\end{equation}

Next we renormalise $\phi^3$ to one loop order.
We will look at the correlation function
\begin{equation}
G(\frac16g\mu^{3-d/2}\phi^3)\equiv\langle\phi(x_1)\phi(x_2)\phi(x_3)\frac16g\mu^{3-d/2}\phi^3(x)\rangle.
\label{phi3greenfn}
\end{equation}
\begin{figure*}[h]
	\centering
	\begin{subfigure}[b]{0.18\textwidth}
	\begin{tikzpicture}
	\begin{feynman}
	\vertex (i);
	\node[crossed dot, right= of i] (a);
	\vertex[above right= of a] (f1);
	\vertex[below right= of a] (f2);
	\diagram*{
		(i) -- (a)
		-- (f1),
		(a) -- (f2)
	};
	\end{feynman}
	\end{tikzpicture}
	\caption{}
	\end{subfigure}
	\begin{subfigure}[b]{0.2\textwidth}
		\begin{tikzpicture}
		\begin{feynman}
		\vertex (i) at (0,0);
		\vertex (a) at (0.7071,-0.7071);
		\node[crossed dot] (c) at (1.7071,-2.4392);
		\vertex (b) at (2.7071,-0.7071);
		\vertex (j) at (1.7071,-3.4392);
		\vertex (k) at (3.414,0);
		\diagram* {
			(i) -- (a)
			-- (b)
			-- (c)
			-- (a),
			(j) -- (c),
			(k) -- (b)
		};
		\end{feynman}
		\end{tikzpicture}
		\caption{}
	\end{subfigure}
	\begin{subfigure}[b]{0.3\textwidth}
		\begin{tikzpicture}
		\begin{feynman}
		\vertex (i);
		\vertex[right= 1cm of i] (a);
		\node[right= of a,crossed dot] (b);
		\vertex[right= 1cm of b] (c);
		\vertex[above right= of c] (f1);
		\vertex[below right= of c] (f2);
		\diagram* {
			(i) -- (a)
			-- [half left] (b)
			-- [half left] (a),
			(b) -- (c)
			-- (f1),
			(c) -- (f2)
		};
		\end{feynman}
		\end{tikzpicture}
		\caption{}
		\label{phi3selfenergy}
	\end{subfigure}
\begin{subfigure}[b]{0.3\textwidth}
	\begin{tikzpicture}
	\begin{feynman}
	\vertex (i);
	\node[right= 1cm of i,crossed dot] (a);
	\vertex[right= of a] (b);
	\vertex[right= 1cm of b] (c);
	\vertex[above right= of c] (f1);
	\vertex[below right= of c] (f2);
	\diagram* {
		(i) -- (a)
		-- [half left] (b)
		-- [half left] (a),
		(b) -- (c)
		-- (f1),
		(c) -- (f2)
	};
	\end{feynman}
	\end{tikzpicture}
	\caption{}
	\label{phi3selfenergy2}
\end{subfigure}
	\begin{subfigure}[b]{0.13\textwidth}
		\begin{tikzpicture}
		\begin{feynman}
		\vertex (i);
		\vertex[right= 1cm of i] (a);
		\vertex[right= 1cm of a] (d);
		\node[above= 0.5cm of a, crossed dot] (b);
		\vertex[above= of b] (c);
		\vertex[above right= of d] (f1);
		\vertex[below right= of d] (f2);
		\diagram*{
			(i) -- (d),
			(a) -- (b)
			-- [half left](c)
			-- [half left](b),
			(d) -- (f1),
			(d) -- (f2) 
		};
		\end{feynman}
		\end{tikzpicture}
		\caption{}
		\label{phi3tadpole}
	\end{subfigure}
	\label{phi3renorm}
	\caption{Renormalisation of $\phi^3$ to one-loop}
\end{figure*}
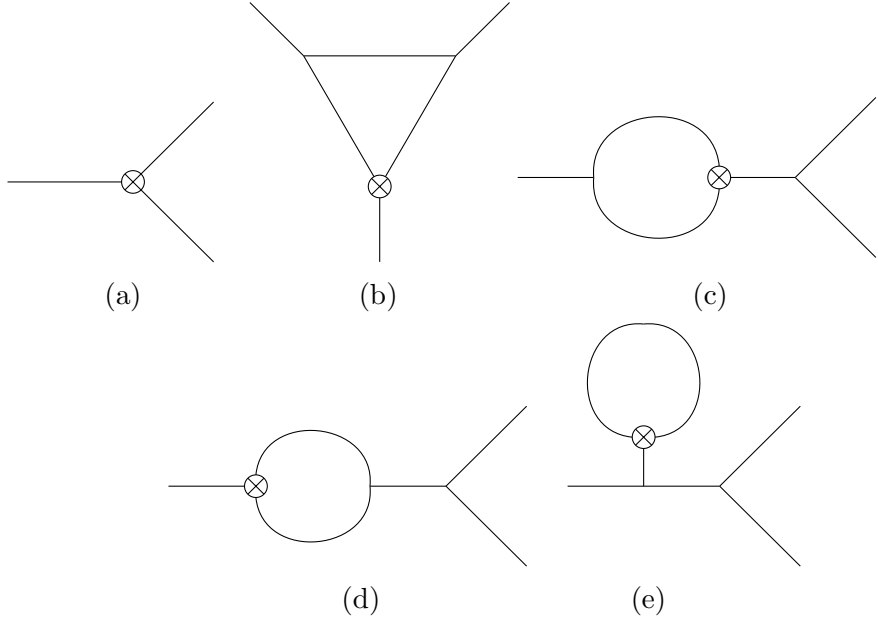
We refer to the five diagrams respectively $G_a$, $G_b$, $G_c$, $G_d$, and $G_e$.
\[\therefore G(\frac16g\mu^{3-d/2}\phi^3)=G_a+3G_b+\sum_{\textrm{3 perms}}\{G_c+G_d+G_e\}.\]
The corresponding momentum space correlation functions, with the composite
operator kept in position space are calculated as follows.
\[G_a=\Delta(p_1)\Delta(p_2)\Delta(p_3).\]
The \(G_b\) calculation is essentially the same as the vertex
renormalisation (bar the \(-i\) that comes with one of the vertices),
i.e., the integral before \(\eqref{vertexcounterterm}\).
\[G_b=\bigg[-\frac{g^3\mu^{3-d/2}}{(4\pi)^3(d-6)}+finite\bigg]\Delta(p_1)\Delta(p_2)\Delta(p_3).\]
Similarly, the \(G_c, G_d\) calculations are reproductions of the
self-energy calculation \(\eqref{massCT}\), giving
\[G_c=\bigg[\frac {ig^3\mu^{3-d/2}}{(4\pi)^3(d-6)}(m^2-(p_2+p_3)^2/6)+finite\bigg]\Delta(p_1)\Delta(p_2)\Delta(p_3)\Delta(p_2+p_3),\]
and
\[G_d=\bigg[\frac {ig^3\mu^{3-d/2}}{(4\pi)^3(d-6)}(m^2-p_1^2/6)+finite\bigg]\Delta(p_1)\Delta(p_2)\Delta(p_3)\Delta(p_2+p_3).\]
Finally, the \(G_e\) calculation reproduces the source term integral
\eqref{deltaj},
\[G_e=\bigg[\frac{g^3\mu^{3-d/2}m^4}{2(4\pi)^3(d-6)}+finite\bigg]\Delta(p_1)\Delta(p_2)\Delta(p_3)\Delta(p_2+p_3)\Delta(p_1+p_2+p_3).\]

To get rid of these poles in the minimal subtraction scheme, we need
counter-terms involving the following correlation functions, (along with the
$G_a$).
To tree level,
\begin{equation}
\begin{aligned}
G(\frac12m^2\phi^2)&=-ig\mu^{3-d/2}m^2\Delta(p_1)\Delta(p_2)\Delta(p_3)\Delta(p_2+p_3)+2 \ perms,\\
G(\phi\partial^2\phi)&=-ig\mu^{3-d/2}[-p_1^2-(p_2+p_3)^2]\Delta(p_1)\Delta(p_2)\Delta(p_3)\Delta(p_2+p_3)+2 \ perms,\\
G(\mu^{d/2-3}\phi)&=-g^2\mu^{3-d/2}\Delta(p_1)\Delta(p_2)\Delta(p_3)\Delta(p_2+p_3)\Delta(p_1+p_2+p_3)+2 \ perms.
\end{aligned}
\end{equation}

Therefore, the finite correlation function corresponding to the correlation function
in \(\eqref{phi3greenfn}\), which we label by
\(G(1/6g\mu^{3-d/2}[\phi^3])\), is given by
\begin{equation}
\boxed{
	\begin{aligned}
	G(\frac16g\mu^{3-d/2}[\phi^3])&=\bigg(1+\frac{3g^2}{(4\pi)^3(d-6)}\bigg)G(\frac16g\mu^{3-d/2}\phi^3)+\frac{2g^2}{(4\pi)^3(d-6)}G(\frac12m^2\phi^2)\\
	&\quad+\frac{g^2}{6(4\pi)^3(d-6)}G(\phi\partial^2\phi)+\frac{g^2}{2(4\pi)^3(d-6)}G((m^4/g)\mu^{d/2-3}\phi)+O(g^3).
	\end{aligned}}
\end{equation}

To compare this with the result we derived in \(\eqref{3-matrix-eq}\)
and in \(\eqref{matrixrow1}\), i.e.,
\begin{equation}
\begin{aligned}
\frac16g_0\phi_0^3&=\Big(1+\frac{\beta/g}{d/2-3}\Big)\frac16g\mu^{3-d/2}[\phi^3]+\frac\delta{d/2-3}\frac12m^2[\phi^2]-\frac\gamma{d/2-3}[E]\\
&\quad-\frac\eta{d/2-3}j\mu^{d/2-3}\phi-\frac{m^4X}g\mu^{d/2-3}\phi+a_4\partial^2\phi+a_6\partial^4\phi+a_7\partial^2[\phi^2].
\end{aligned}
\label{matrixresult}
\end{equation}
We have to \(O(g^2)\) the coefficients of the first four terms,
\begin{equation}
\begin{aligned}
\beta=-\frac{3g^3}{4(4\pi)^3};\quad\delta=-\frac{5g^2}{6(4\pi)^3};\quad\gamma=\frac{g^2}{12(4\pi)^3};\\
\eta=\frac{g^2}{2(4\pi)^3}\Big(\frac16+\frac{m^4}{jg}\Big);\quad X=-\frac{g^2}{4(4\pi)^3(d/2-3)}.
\end{aligned}
\end{equation}
It is apparent from this comparison that for this to be consistent, we
must have
\[a_4=a_6=a_7=0+O(g^3).\]
\[\therefore \zeta_4=\zeta_6=\zeta_7=0+O(g^3).\]

Some algebra is necessary before \eqref{matrixresult} can be
compared with Feynman diagram computation.
\begin{equation}
\begin{aligned}
\frac16(g+\delta g)\mu^{3-d/2}\phi^3&=\bigg(1-\frac{3g^2}{4(4\pi)^3(d/2-3)}\bigg)\frac16g\mu^{3-d/2}[\phi^3]-\frac{5g^2}{6(4\pi)^3(d/2-3)}\frac12m^2\phi^2\\
&\quad-\frac{g^2}{12(4\pi)^3(d/2-3)}\{\phi\partial^2\phi+m^2\phi^2+\frac12g\mu^{3-d/2}\phi^3-j\mu^{d/2-3}\phi\}\\
&\quad -\frac{g^2}{12(4\pi)^3(d/2-3)}j\mu^{d/2-3}\phi+\bigg(\frac14-\frac12\bigg)\frac{g^2}{(4\pi)^3(d/2-3)}\frac{m^4}g\mu^{d/2-3}\phi+O(g^3).
\end{aligned}
\end{equation}
Here we have substituted for the renormalised term \([\phi^2]\) with
\(\phi^2\), since we needed to keep only \(O(1)\) terms, similarly in
\(E_0=[E]\), counterterms for all the terms are ignored, as they would
be \(O(g^2)\). Substituting for \(\delta g\) using
\eqref{vertexcounterterm},
\begin{equation}
\begin{aligned}
\frac16g\mu^{3-d/2}[\phi^3]&=\bigg(1+\frac{g^2}{(4\pi)^3(d-6)}+\frac{g^2}{4(4\pi)^3(d/2-3)}\bigg)\bigg(1-\frac{3g^2}{4(4\pi)^3(d/2-3)}\bigg)^{-1}\frac16g\mu^{3-d/2}\phi^3\\
&\quad+\frac{g^2}{(4\pi)^3(d/2-3)}\frac12m^2\phi^2 +\frac{g^2}{12(4\pi)^3(d/2-3)}\phi\partial^2\phi\\
&\quad +\frac14\frac{g^2}{(4\pi)^3(d/2-3)}\frac{m^4}g\mu^{d/2-3}\phi+O(g^3)\\
&=\bigg(1+\frac{3g^2}{2(4\pi)^3(d/2-3)}\bigg)\frac16g\mu^{3-d/2}\phi^3+\frac{g^2}{(4\pi)^3(d/2-3)}\frac12m^2\phi^2 \\
&\quad+\frac{g^2}{12(4\pi)^3(d/2-3)}\phi\partial^2\phi +\frac14\frac{g^2}{(4\pi)^3(d/2-3)}\frac{m^4}g\mu^{d/2-3}\phi+O(g^3)
\end{aligned}
\end{equation}
That matches for the terms without total derivative operators.

\bibliography{RG,Background}

\begin{thebibliography}{10}

\bibitem{brown_dimensional_1980}
Lowell~S Brown.
\newblock Dimensional regularization of composite operators in scalar field
  theory.
\newblock {\em Annals of Physics}, 126(1):135--153, April 1980.

\bibitem{polchinski_renormalization_1984}
Joseph Polchinski.
\newblock Renormalization and effective lagrangians.
\newblock {\em Nuclear Physics B}, 231(2):269--295, January 1984.

\bibitem{zimmermann_local}
Wolfhart Zimmermann.
\newblock Local operator products and renormalization in quantum field theory.
\newblock In Stanley Deser, Marc Grisaru, and Hugh Pendleton, editors, {\em
  Lectures On Elementary Particles and Quantum Field Theory}, volume~1. M.I.T.
  Press, 1970.

\bibitem{Breitenlohner:1977hr}
P.~Breitenlohner and D.~Maison.
\newblock {Dimensional Renormalization and the Action Principle}.
\newblock {\em Commun. Math. Phys.}, 52:11--38, 1977.

\bibitem{collins_normal_1975}
J.~C. Collins.
\newblock Normal products in dimensional regularization.
\newblock {\em Nuclear Physics B}, 92(4):477--506, January 1975.

\bibitem{KlubergStern:1974rs}
H.~Kluberg-Stern and J.~B. Zuber.
\newblock Ward identities and some clues to the renormalization of
  gauge-invariant operators.
\newblock {\em Phys. Rev. D}, 12:467--481, Jul 1975.

\bibitem{Zee:1983mj}
A.~Zee.
\newblock {Einstein Gravity Emerging From Quantum Weyl Gravity}.
\newblock {\em Annals Phys.}, 151:431, 1983.

\bibitem{Freeman:1983cx}
M.~D. Freeman.
\newblock {The Renormalization of Nonabelian Gauge Theories in Curved
  Space-time}.
\newblock {\em Annals Phys.}, 153:339, 1984.

\bibitem{Dudal:2008tg}
D.~Dudal, S.~P. Sorella, N.~Vandersickel, and H.~Verschelde.
\newblock {A Purely algebraic construction of a gauge and renormalization group
  invariant scalar glueball operator}.
\newblock {\em Eur. Phys. J. C}, 64:147--159, 2009, 0812.2401.

\bibitem{Spiridonov:1984br}
V.~P. Spiridonov.
\newblock {Anomalous Dimension of $G^2_{\mu\nu}$ and $\beta$ Function}.
\newblock 1984.

\bibitem{Hathrell:1981gz}
S.~J. Hathrell.
\newblock {Trace Anomalies and {QED} in Curved Space}.
\newblock {\em Annals Phys.}, 142:34, 1982.

\bibitem{Osborn:1987au}
H.~Osborn.
\newblock {Renormalization and Composite Operators in Nonlinear $\sigma$
  Models}.
\newblock {\em Nucl. Phys. B}, 294:595--620, 1987.

\bibitem{Benedetti:2020yvb}
Dario Benedetti, Razvan Gurau, and Kenta Suzuki.
\newblock {Conformal symmetry and composite operators in the $O(N)^{3}$ tensor
  field theory}.
\newblock {\em JHEP}, 06:113, 2020, 2002.07652.

\bibitem{Lalak:2018bow}
Zygmunt Lalak and Pawel Olszewski.
\newblock {Vanishing trace anomaly in flat spacetime}.
\newblock {\em Phys. Rev. D}, 98(8):085001, 2018, 1807.09296.

\bibitem{Bellucci:2003ud}
S.~Bellucci, I.~L. Buchbinder, and V.~A. Krykhtin.
\newblock {Renormalization of the energy momentum tensor in noncommutative
  scalar field theory}.
\newblock {\em Nucl. Phys. B}, 665:402--424, 2003, hep-th/0303186.

\bibitem{Mauri:2021ili}
Achille Mauri and Mikhail~I. Katsnelson.
\newblock {Scale without conformal invariance in membrane theory}.
\newblock 4 2021, 2104.06859.

\bibitem{callan_new_1970}
Curtis~G Callan, Sidney Coleman, and Roman Jackiw.
\newblock A new improved energy-momentum tensor.
\newblock {\em Annals of Physics}, 59(1):42--73, July 1970.

\bibitem{Lowenstein:1971vf}
J.~H. Lowenstein.
\newblock {Normal product quantization of currents in Lagrangian field theory}.
\newblock {\em Phys. Rev. D}, 4:2281--2290, 1971.

\bibitem{Freedman:1974gs}
Daniel~Z. Freedman, Ivan~J. Muzinich, and Erick~J. Weinberg.
\newblock {On the Energy-Momentum Tensor in Gauge Field Theories}.
\newblock {\em Annals Phys.}, 87:95, 1974.

\bibitem{Freedman:1974ze}
Daniel~Z. Freedman and Erick~J. Weinberg.
\newblock {The Energy-Momentum Tensor in Scalar and Gauge Field Theories}.
\newblock {\em Annals Phys.}, 87:354, 1974.

\bibitem{collins_renormalization_1976}
J.~C. Collins.
\newblock Renormalization of the energy-momentum tensor in
  ${\ensuremath{\varphi}}^{4}$ theory.
\newblock {\em Phys. Rev. D}, 14:1965--1976, Oct 1976.

\bibitem{Polchinski:1987dy}
Joseph Polchinski.
\newblock {Scale and Conformal Invariance in Quantum Field Theory}.
\newblock {\em Nucl. Phys. B}, 303:226--236, 1988.

\bibitem{Coleman:1970je}
Sidney~R. Coleman and Roman Jackiw.
\newblock {Why dilatation generators do not generate dilatations?}
\newblock {\em Annals Phys.}, 67:552--598, 1971.

\bibitem{Christensen:1977jc}
S.~M. Christensen and S.~A. Fulling.
\newblock {Trace Anomalies and the Hawking Effect}.
\newblock {\em Phys. Rev. D}, 15:2088--2104, 1977.

\bibitem{Bibilashvili:1992hx}
T.~Bibilashvili and I.~Paziashvili.
\newblock {Real time finite temperature green functions for nonuniform
  nonequilibrium media}.
\newblock {\em Annals Phys.}, 220:134--155, 1992.

\bibitem{Joglekar:1988uc}
Satish~D. Joglekar and Anuradha Misra.
\newblock {A Uniqueness Theorem Regarding $\theta (\mu \nu$) in Scalar
  Theories}.
\newblock {\em Annals Phys.}, 185:231--240, 1988.

\bibitem{Cardy:1974af}
John~L. Cardy.
\newblock {Regge Behavior in as Asymptotically Free Field Theory}.
\newblock {\em Phys. Lett. B}, 53:355--358, 1974.

\bibitem{Lovelace:1974mi}
C.~Lovelace.
\newblock {Regge Behavior Under Asymptotic Freedom}.
\newblock {\em Phys. Lett. B}, 55:187--191, 1975.

\bibitem{Cardy:1975fz}
John~L. Cardy.
\newblock {High-Energy Behavior in phi**3 Theory in Six-Dimensions}.
\newblock {\em Nucl. Phys. B}, 93:525--546, 1975.

\bibitem{Ma:1975vn}
Ernest Ma.
\newblock {Asymptotic Freedom and a Quark Model in Six-Dimensions}.
\newblock {\em Prog. Theor. Phys.}, 54:1828, 1975.

\bibitem{Cornwall:1995dr}
J.~M. Cornwall and D.~A. Morris.
\newblock {Toy models of nonperturbative asymptotic freedom in phi**3 in
  six-dimensions}.
\newblock {\em Phys. Rev. D}, 52:6074--6086, 1995, hep-ph/9506293.

\bibitem{collins_1984}
John~C. Collins.
\newblock {\em Renormalization: An Introduction to Renormalization, the
  Renormalization Group and the Operator-Product Expansion}.
\newblock Cambridge Monographs on Mathematical Physics. Cambridge University
  Press, 1984.

\bibitem{srednicki_2007}
Mark Srednicki.
\newblock {\em Quantum Field Theory}.
\newblock Cambridge University Press, 2007.

\bibitem{Toms:1982af}
David~J. Toms.
\newblock {Renormalization of Interacting Scalar Field Theories in Curved
  Space-time}.
\newblock {\em Phys. Rev. D}, 26:2713, 1982.

\bibitem{fisher_yang-lee_1978}
Michael~E. Fisher.
\newblock Yang-lee edge singularity and ${\ensuremath{\phi}}^{3}$ field theory.
\newblock {\em Phys. Rev. Lett.}, 40:1610--1613, Jun 1978.

\bibitem{Gracey:2015tta}
J.~A. Gracey.
\newblock {Four loop renormalization of $\phi^3$ theory in six dimensions}.
\newblock {\em Phys. Rev. D}, 92(2):025012, 2015, 1506.03357.

\bibitem{deAlcantaraBonfim:1981sy}
O.~F. de~Alcantara~Bonfim, J.~E. Kirkham, and A.~J. McKane.
\newblock {Critical Exponents for the Percolation Problem and the Yang-lee Edge
  Singularity}.
\newblock {\em J. Phys. A}, 14:2391, 1981.

\bibitem{An:2016lni}
Xin An, David Mesterh\'azy, and Mikhail~A. Stephanov.
\newblock {Functional renormalization group approach to the Yang-Lee edge
  singularity}.
\newblock {\em JHEP}, 07:041, 2016, 1605.06039.

\bibitem{Fei:2014yja}
Lin Fei, Simone Giombi, and Igor~R. Klebanov.
\newblock {Critical $O(N)$ models in $6-\epsilon$ dimensions}.
\newblock {\em Phys. Rev. D}, 90(2):025018, 2014, 1404.1094.

\bibitem{Mack:1973kaa}
G.~Mack.
\newblock {Conformal invariance and short distance behavior in quantum field
  theory}.
\newblock {\em Lect. Notes Phys.}, 17:300--334, 1973.

\bibitem{Gliozzi:2014jsa}
Ferdinando Gliozzi and Antonio Rago.
\newblock {Critical exponents of the 3d Ising and related models from Conformal
  Bootstrap}.
\newblock {\em JHEP}, 10:042, 2014, 1403.6003.

\bibitem{macfarlane_3_1974}
A.~J. Macfarlane and G.~Woo.
\newblock ${\ensuremath{\Phi}}^{3}$ theory in six dimensions and the
  renormalization group.
\newblock {\em Nuclear Physics B}, 77(1):91--108, July 1974.

\bibitem{t_hooft_dimensional_1973}
G.~'t~Hooft.
\newblock Dimensional regularization and the renormalization group.
\newblock {\em Nuclear Physics B}, 61:455--468, September 1973.

\bibitem{collins_new_1974}
J.~C. Collins and A.~J. Macfarlane.
\newblock New methods for the renormalization group.
\newblock {\em Phys. Rev. D}, 10(4):1201--1212, August 1974.

\bibitem{t_hooft_regularization_1972}
G.~'t~Hooft and M.~Veltman.
\newblock Regularization and renormalization of gauge fields.
\newblock {\em Nuclear Physics B}, 44(1):189--213, July 1972.

\bibitem{brown_1992}
Lowell~S. Brown.
\newblock {\em Quantum Field Theory}.
\newblock Cambridge University Press, 1992.

\bibitem{Dutta:2020vqo}
S.~Dutta, B.~Sathiapalan, and H.~Sonoda.
\newblock {Wilson action for the $O(N)$ model}.
\newblock {\em Nucl. Phys. B}, 956:115022, 2020, 2003.02773.

\bibitem{wilson_quantum_1973}
Kenneth~G. Wilson.
\newblock Quantum {Field} - {Theory} {Models} in {Less} {Than} 4 {Dimensions}.
\newblock {\em Phys. Rev. D}, 7(10):2911--2926, May 1973.

\end{thebibliography}
\bibliographystyle{hunsrt.bst}

\end{document}